\documentclass[usenatbib,useAMS,manuscript]{mn2e}
\usepackage{amssymb}
\usepackage{graphicx}
\usepackage{multirow}

\title
[Mass profiles of elliptical galaxies]
{Modelling mass distribution in elliptical galaxies: mass profiles
 and their correlation with velocity dispersion profiles} 

\author[Chae, Bernardi \& Kravtsov]
{Kyu-Hyun~Chae$^{1\star}$, Mariangela Bernardi$^{2}$ and 
Andrey V. Kravtsov$^{3,4}$ \\
\\
$^1$Sejong University, Department of Astronomy and Space Science, 
 98 Gunja-dong, Gwangjin-Gu, Seoul 143-747, Republic of Korea\\
$^{2}$Department of Physics and Astronomy, University of 
       Pennsylvania, 209 South 33rd Street, Philadelphia, PA 19104, USA \\
$^{3}$Kavli Institute for Cosmological Physics, 
 5640 South Ellis Avenue, The University of Chicago, Chicago, IL 60637, USA \\
$^{4}$Department of Astronomy and Astrophysics, 
 5640 South Ellis Avenue, The University of Chicago, Chicago, IL 60637, USA \\
$^\star$chae@sejong.ac.kr}
\date{
Accepted ........
Received .......;
in original form ......}

\pubyear{2013}

\begin{document}

\maketitle

\begin{abstract}
We assemble a statistical set of global mass models for
$\sim$~2,000 nearly spherical Sloan Digital Sky Survey (SDSS) galaxies
at a mean redshift of $\langle z\rangle=0.12$ based on
their aperture velocity dispersions and newly derived 
 luminosity profiles in conjunction with published velocity dispersion
profiles and empirical properties and relations of galaxy and halo parameters.
When two-component (i.e.\ stellar plus dark) mass models are fitted to the 
SDSS aperture velocity dispersions, the predicted velocity dispersion profile 
(VP) slopes within the effective (i.e.\ projected half-light) radius 
$R_{\rm eff}$ match well the distribution in observed elliptical galaxies. 
 From a number of input variations the models exhibit 
 for the radial range $0.1 R_{\rm eff}< r < R_{\rm eff}$ a tight correlation 
$\langle\gamma_{\rm e}\rangle=(1.865\pm 0.008)+(-4.93\pm 0.15)\langle\eta\rangle$
where $\langle\gamma_{\rm e}\rangle$ is the mean slope absolute value of the 
total mass density and $\langle\eta\rangle$ is the mean slope of the velocity 
dispersion profile, which 
leads to a super-isothermal $\langle\gamma_{\rm e}\rangle = 2.15\pm 0.04$ 
for $\langle\eta\rangle =-0.058\pm 0.008$ in observed elliptical galaxies.
 Furthermore, the successful two-component models appear to imply a typical 
slope curvature pattern in the total mass profile because for the observed 
steep luminosity (stellar mass) profile and the weak lensing inferred halo
 profile at large radii a total mass profile with monotonically varying slope 
would require too high DM density in the optical region giving rise to too 
large aperture velocity dispersion and too shallow VP.
\end{abstract}


\maketitle

\section{Introduction}

Observed galaxies exhibit substantial diversity in their morphological 
appearances and structure. Early-type, namely elliptical and lenticular, 
galaxies (ETGs) are more massive on average among galaxy types and are thought 
to originate from mergers of late-type galaxies and smaller ETGs under the 
paradigm of hierarchical galaxy formation and evolution 
(e.g., \citealt{PS74,WR}). The empirical properties of ETGs hold important 
clues to the physical processes shaping properties of the most massive galaxies.
 ETGs also play a crucial role in modern cosmology including dark energy 
phenomenology. Owing to their large masses and high central densities ETGs 
dominate observed population of strong lenses and are also important for weak 
lensing which provide both geometric (through cosmological distances) and
dynamical (through growth of structures) probe of dark energy (\citealt{FTH}). 
Massive ETGs, or similarly luminous red galaxies by colour selection, are also 
excellent targets of the baryonic acoustic oscillation studies that provide an 
independent cosmological probe of the distance scale in the universe 
(e.g., \citealt{Eis,Per,San}). 

For both astrophysical and cosmological studies involving ETGs, their
 density structure, in particular the radial profile, is of great 
importance for several reasons. 
First of all, as the radial mass profile encodes the combined distribution
of luminous and dark matter, it is a key quantity to constrain galaxy 
formation and evolution models. An important question is whether there 
is a simple natural attractor of radial density profile such as the 
Navarro-Frenk-White (NFW) profile (e.g., \citealt{NFW,LP})
 or the isothermal profile (e.g., \citealt{LB,Ger,Chae11}). 
Secondly, because there is no empirical method 
to directly measure dark matter (DM) distribution in a galaxy,
 accurate determination of the total mass profile is the crucial first 
step toward the inference of DM distribution (e.g., \citealt{New}). 
Finally, the radial mass profile is a crucial factor in gravitational 
lensing (both strong and weak). Strong lensing image flux ratios depend 
largely on the radial density slope at the Einstein radius. 
 Weak lensing by individual haloes and large-scale structures depend on the
global radial profile of galaxies and clusters.

The radial total mass profile of a galaxy can be constrained using 
dynamic probes such as stellar or gas dynamics, or gravitational lensing.  
For nearby ETGs stellar dynamics has been the most 
fruitful probe because detailed stellar kinematic data as well as photometric
data are available. Detailed dynamical models based on the observed luminosity 
distributions and stellar velocity moments have been constructed for tens to 
hundreds ETGs (e.g., \citealt{Ger,Tho,Cap13a}). For nearby $X$-ray bright 
massive ellipticals, hydrostatic equilibrium equation has 
been used to infer mass profiles (e.g., \citealt{Das}).   
The inferred mass profiles from these studies in general exhibit significant 
variation of the logarithmic slope with radius, although the density profiles 
are close to isothermal around the half-light radius of stellar distribution. 
Nevertheless, the sample sizes of dynamical and lensing studies of mass 
distribution in the ETGs have been fairly small to allow systematic study of 
profile variation in a wide radial range.

Strong lensing studies have been used to constrain the total mass profile of
 distant galaxies at various redshifts (e.g., \citealt{RKK,Bar,Ruf,Bol}). 
Strong lensing effectively constrains the density slope (absolute value)
$\gamma (\equiv -d\ln\rho/d\ln r)$ or,
 more precisely, the slope of the projected 2-dimensional density 
$\gamma_{\rm 2D}$, at the Einstein radius $R_{\rm Ein}$, which increases with 
redshift at fixed $R_{\rm eff}$: e.g., 
$R_{\rm Ein} \sim 0.5R_{\rm eff}$ at $z = 0.1$ but 
$R_{\rm Ein} \sim R_{\rm eff}$ at $z = 0.6$ (\citealt{Bol}).
For systematic analyses of samples of tens of strong lens systems 
constant-$\gamma$ models have usually been adopted and the constrained values 
of $\gamma$ have mean slopes close to the isothermal value 
 $|\langle\gamma\rangle - 2|<0.1$ for 
$0.2\lesssim z\lesssim 1$ (e.g., \citealt{RKK,Bar}). 
However, recently it was noticed that the value of 
$\gamma$ appeared to evolve significantly with $z$ based on combined samples 
that contain both low $z \sim 0.1$ lenses and intermediate $z \sim 0.6$ lenses
 (\citealt{Bol,Ruf}). Assuming $\gamma$ was independent of radius in the optical
 region, the apparent redshift evolution of $\gamma$ was interpreted in terms of
 galaxy evolution. Notably, the apparent evolution implies 
$\langle\gamma\rangle \approx 2.3$ at $z=0$ (\citealt{Bol}). 
 Finally,  stellar dynamics, strong and weak lensing were combined to infer the
 total mass profile for a large radial range for several individual systems 
(\citealt{New}) or stacked data (\citealt{Gav}). 
The former work focused on galaxy clusters, while the latter study used only the
de~Vaucouleurs stellar mass profile and the NFW DM profile without taking into 
account halo contraction effects by baryonic physics 
(e.g., \citealt{Blu,Gne,SM,Gne11}).
 
 From recent dynamical analyses of ETGs it is not clear  whether the 
isothermal model is a good approximation to the average profile for some radial
range or whether the average profile is systematically varying with redshift 
or some other parameter. 
Here we investigate galaxy mass profile through Jeans dynamical analyses
 based on Sloan Digital Sky Survey (SDSS; \citealt{York}) galactic luminosity 
distributions and aperture velocity dispersions in conjunction with published 
velocity dispersion profiles and empirical properties and relations for ETGs 
and their haloes. SDSS photometric data and aperture velocity dispersions 
have been useful in addressing several issues including estimating dynamical 
masses and dark matter contents within $R_{\rm eff}$ (e.g., \citealt{Pad,Tor12}),
 testing halo contraction/expansion models and  
stellar initial mass functions (IMFs) (\citealt{Dut11,Dut13,DT13}).
These studies confirm that non-negligible amount 
of dark matter is generally required within $R_{\rm eff}$ 
 and suggest that the DM fraction within $R_{\rm eff}$ is correlated with
 properties such as surface brightness and $R_{\rm eff}$  and 
 galaxies require non-universal IMFs or/and 
non-universal dark halo response to galaxy formation. 
Our present study of SDSS ETGs has the following aspects. 
First, we select nearly spherical and disk-less galaxies
for our analyses by the spherical Jeans equation so that any possible
systematic errors due to non-spherical shapes, which were ignored in all 
previous analyses, can be minimized. Second, the main focus of our work is 
galaxy total mass profile and for that purpose we use an empirical distribution
 of velocity dispersion profiles (VPs) within $R_{\rm eff}$. Third, we use newly
measured luminosity profile parameters, Sersic index $n$ and
effective radius $R_{\rm eff}$. Fourth, we take into account recently published
systematic IMF variation in early-type galaxies. Finally, our analyses are
based on real data and auxiliary empirical inputs excluding simulation or
theoretical inputs. Our present work has a different focus and is an improvement
with updated and more reliable inputs compared with \cite{Chae12} which 
considered VPs but focused on DM profiles with inputs from N-body simulations.

For our Jeans analyses we assume parametric functional forms for the total mass
profile. When a model is fitted to the aperture velocity dispersion, it predicts
a velocity dispersion profile. We then compare the predicted distribution of
VPs with the empirical distribution  to test the model.
 Our successful models are two-component models in which stellar mass 
distribution is assumed to follow the photometrically derived luminosity 
profile and the coupled DM distribution is assumed to follow the gNFW or 
Einasto model, which are motivated from cosmological hydrodynamic simulations 
of galaxy formation, with the constraint that they match weak lensing 
observations at large radii.
 Our model set is at least an order
 of magnitude larger than any previous  set of empirical 
(as opposed to simulation or semi-empirical) models for ETGs. In addition, 
it covers a large radial range extending from $\sim 0.1 R_{\rm eff}$
 to the halo virial radius.

The outline of this paper is as follows. In section~2 we 
 describe our data and other empirical results from the literature 
that are needed for our analyses.
In section~3 we describe mass model parametrisation and method of Jeans 
dynamical analyses. In section~4 we describe statistical results for mass 
profiles based on standard empirical inputs and then in section~5 we consider
systematic variations of the statistical results. In section~6 we compare our
results with recent results based on SDSS photometric data and aperture
velocity dispersions. 
In sections~7 \& 8 we discuss implications of our results for galaxy formation
 and strong lensing. We conclude in section~9.

\section{Galaxy data and empirical relations}

We carefully define a sample of nearly spherical (projected 
minor-to-major axis ratio $b/a > 0.85$) and disk-less
(disk mass is within the measurement error of total mass) galaxies
with mean redshift $\langle z \rangle \approx 0.12$ selected from the 
 SDSS (\citealt{Ber,Ber12,Mee}). 
This selection excludes all lenticular and later-type galaxies, 
while retaining galaxies spanning two orders of magnitude in stellar mass 
($10 \lesssim \log_{10}(M_{\star}/{\rm M}_{\odot}) \lesssim 12$), 
and was chosen to minimize biases in analyses
 based on the spherical Jeans equation (see section~3).
Therefore, our resulting mass profiles will apply strictly to nearly spherical 
galaxies only.
However, we will also consider modelling a general sample of ETGs
to estimate any systematic difference in mass profile between our nearly
spherical sample and a general sample.

Because each galaxy in our sample has a nearly spherical component only, 
its luminosity profile can generally be well described by a single-component
 \cite{Ser} profile, except possibly for the sub-kiloparsec 
central region (see below).
Each galaxy has the surface brightness S\'{e}rsic-fit effective radius 
$R_{\rm eff}$, the S\'{e}rsic index $n$, the luminosity-weighted line-of-sight 
velocity dispersion (LOSVD) 
within the SDSS aperture of radius 1.5~arcsec $\sigma_{\rm ap}$ and 
 the total stellar mass $M_{\star}^{\rm Ch}$ converted from the measured luminosity
using the Chabrier(\citealt{Chab}) stellar initial mass function (IMF) or 
equivalently mass-to-light ($M/L$) ratio. 
Recent observational studies have found that IMFs 
of ETGs are not universal but show galaxy-to-galaxy scatter and systematic 
trends with some parameters such as stellar 
 velocity dispersion (VD) $\sigma$. 
 To encompass the current likely range we use three independently inferred 
$\sigma$-dependent IMFs  (e.g., \citealt{Cap13b,CvD,Tor}; see section~2.3).

It is well-known  that the central sub-kiloparsec region of 
ETG light profile exhibits large variation giving rise to either the
central missing light due to the shallower core or the central extra light due 
to the steeper power-law profile compared with 
the main body profile extrapolation (e.g., \citealt{Fab,Hyd,Kor,Gla}). 
This central extra or missing light can have non-negligible effects for 
small $R_{\rm eff}$ galaxies. However, SDSS photometric data do not have 
enough resolution for the central regions of our $\langle z\rangle\approx 0.12$
galaxies.
 To allow for the observed deviations in the central regions from the overall 
S\'{e}rsic profiles for ETGs we use the central profile properties, i.e.\ the 
break radius and the slope at a fiducial central radius as a function of 
stellar mass, extracted from published data 
including the {\it Hubble Space Telescope} data  (see section~2.8).

Each galaxy is further assigned its halo mass $M_{200}$ within the virial 
radius $r_{200}$, which is defined to be the radius of the sphere within which 
the mean density is 200 times the cosmic mean matter density 
(equation~6 of \citealt{Man08}),
\begin{equation}
  r_{200} = \frac{251.6}{(1+z)} 
         \left(\frac{M_{200}}{10^{12} h^{-1}{\rm M}_{\odot}}\right)^{1/3} 
 h^{-1}~{\rm kpc}  
\label{eq:rvir}
\end{equation}
assuming a flat $\Lambda$-dominated Universe with $\Omega_{\rm m0}=0.27$ 
and  $h=H_0/100~{\rm km}~{\rm s}^{-1}=0.7$ (\citealt{Kom}). 
For this we use the empirical $M_{\star}$-$M_{200}$ relations from SDSS satellite
 kinematics (\citealt{More}) and weak lensing (\citealt{Sch,Man06}), as well as
 the abundance matching relation (\citealt{Chae12}) to encompass the current 
likely range (see section~2.4). 

The halo assigned to each galaxy is then assigned a mass profile at large radii
($r > 0.2 r_{200}$) using the weak lensing observations of the SDSS galaxies, 
which not only show that the outer halo profile 
can be well described by the NFW profile but also give the mean halo 
mass ($M_{200}$)-NFW concentration ($c_{\rm NFW}$) relation derived from stacked 
profiles (\citealt{Man08}). We use an intrinsic scatter of $0.1$~dex and allow 
$\pm 30\%$ of systematic errors.

To constrain the mass profile in the optical region 
  of each nearly spherical galaxy we need not only the
 aperture VD $\sigma_{\rm ap}$ but also the VD profile. 
We do not have a measured VD profile for each galaxy in our sample. 
However, we can use an empirical statistical distribution of VD profiles from 
the literature (\citealt{Cap06,Meh}) as described in section~2.6.

In the following subsections we describe in detail our data and all empirical 
results from the literature that are needed for our analyses.

\subsection{Luminosity (stellar mass) profiles}

To describe the observed light distribution of an ETG we use the 
S\'{e}rsic model or a two-component (`SerExp') model that contains a 
S\'{e}rsic bulge and an exponential disk. 
From the light distribution we derive the volume distribution of 
stellar mass using empirically derived stellar IMFs (see below section~2.3).
 
The S\'{e}rsic surface brightness distribution on the sky given by
\begin{equation}
I(X) \propto \exp\left( -b_n X^{1/n} \right),
\label{eq:ser} 
\end{equation}
with $X=R/R_{\rm eff}$ ($R$ being the two dimensional radius) and 
$b_n = 2n-1/3+0.009876/n$ (for $0.5 < n < 10$) (\citealt{PS}) allows a more 
accurate description of a dispersion-supported system than the traditional 
de~Vaucouleurs model which is the special case of $n=4$.  

More recently it has been recognized that most ETGs are better described
by a compound model composed of two distinct components, namely a bulge and
 a disk, although the bulge is dominating (\citealt{Ber12,Mee}).
 When a two-component system with the non-negligible disk is fitted by a 
single S\'{e}rsic model, the resulting parameters, in particular galaxy size 
$R_{\rm eff}$, can be biased compared with the actual bulge parameters 
(\citealt{Ber12,Mee}). The bias in $R_{\rm eff}$ is less than 0.05~dex 
for stellar mass $\lesssim 10^{11.5}{\rm M}_\odot$ but can be as large as 
0.08~dex near the high mass tail (\citealt{Ber12}).  

 For our Jeans analyses we select nearly spherical and disk-less systems
 that are well-described by the single-component S\'{e}rsic profiles (see, 
however, section~2.8 for the profiles of the central sub-kpc regions). 
However, we also consider modelling general ETGs to estimate possible systematic
difference of our results from those for general ETGs.
In this case we use the single S\'{e}rsic fit to the total light 
distribution.  When we use the single S\'{e}rsic fit for a galaxy having 
a non-negligible disk, 
we are using a biased $R_{\rm eff}$ as mentioned above.

\subsection{SDSS samples of ETGs}

We select a sample of 28,259 ETGs from the SDSS main galaxy sample 
satisfying light concentration $C_r > 2.6$ and $n_{\rm bulge} > 2.5$ where 
$C_r$ is the ratio of the radius containing 90 percent of the Petrosian 
luminosity in the $r$-band to that containing 50 percent (\citealt{Ber}) 
and $n_{\rm bulge}$ is the bulge S\'{e}rsic index 
in the SerExp fit of the total light distribution.
  A subsample of 20,210 ETGs satisfies $C_r > 2.86$.
Each galaxy in the sample has aperture velocity dispersion 
$\sigma_{\rm ap}$ and ellipticity $\varepsilon$ 
($\equiv 1-b/a$ with $b/a$ being the projected minor-to-major axis ratio) as 
well as fitted parameters of both the S\'{e}rsic and the SerExp models along 
with corresponding stellar masses based on the Chabrier IMF. 
The S\'{e}rsic-fit parameters are $M_{\star}^{\rm Ch}$ 
(total stellar mass based on the Chabrier IMF), $R_{\rm eff}$ (effective radius) 
and $n$ (S\'{e}rsic-fit index). The SerExp-fit parameters are 
$M_{{\rm bulge,}\star}^{\rm Ch}$ (bulge stellar mass based on the Chabrier IMF), 
$R_{\rm bulge,eff}$ (bulge effective radius), $n_{\rm bulge}$ (bulge S\'{e}rsic-fit 
index) as well as $M_{\star}^{\rm Ch}$ and $R_{\rm eff}$ for the total (bulge 
plus exponential) profile.

\begin{figure*} %1
\begin{center}
\setlength{\unitlength}{1cm}
\begin{picture}(16,8)(0,0)
\put(-0.5,10.5){\includegraphics{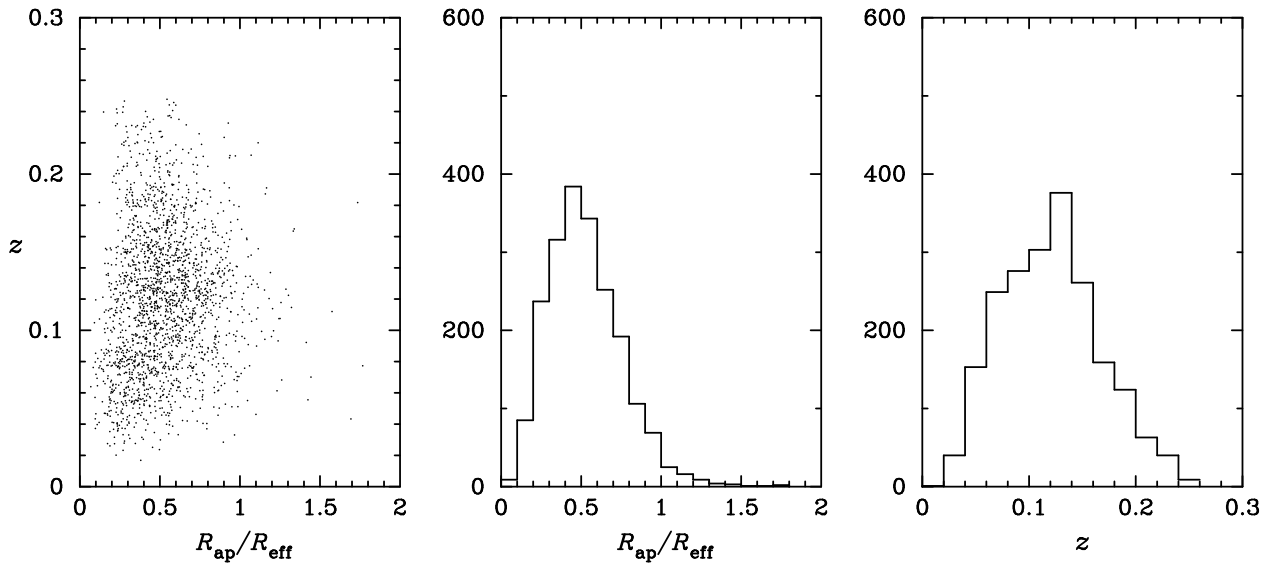}}
\end{picture}
\caption{
Distribution of $R_{\rm ap}/R_{\rm eff}$ (the ratio of the physical aperture 
radius to the effective radius) and $z$ (redshift) in our standard sample of 
2,054 nearly spherical and disk-less SDSS galaxies.
}
\label{Rz}
\end{center}
\end{figure*}

Based on the observed light distributions and the model fitted parameters
 we select a  $C_r > 2.86$ subsample of 2,054 nearly spherical and 
disk-less ellipticals using the criteria 
$\varepsilon < 0.15$ (single S\'{e}rsic-fit 
ellipticity) and $\log_{10}(R_{\rm eff}/R_{\rm bulge,eff}) < 0.19$
(the sum of the estimated measurement errors for $R_{\rm eff}$ in the
single S\'{e}rsic fit and $R_{\rm bulge,eff}$ in the SerExp fit).  
Galaxies in this 
sample have the following properties 
$\log_{10}(M_\star^{\rm Ch}/M_{{\rm bulge,}\star}^{\rm Ch}) = 0.064\pm 0.049$,
 $\log_{10} (R_{\rm eff}/R_{\rm bulge,eff}) = 0.062\pm 0.085$ and 
$\varepsilon = 0.089\pm 0.039$. 
The distribution of redshifts and the physical aperture to effective radius
ratios $R_{\rm ap}/R_{\rm eff}$ can be found in Fig.~\ref{Rz}. The redshift
distribution has a mean of $\langle z\rangle \approx 0.12$ with a 
root-mean-square (RMS) dispersion of $s_z \approx 0.05$ while the 
$R_{\rm ap}/R_{\rm eff}$ distribution has  
$\langle R_{\rm ap}/R_{\rm eff}\rangle \approx 0.53$
and $s_{R_{\rm ap}/R_{\rm eff}} \approx 0.23$.
Another important feature of this sample is that galaxies have higher 
  S\'{e}rsic indices compared with other ETG samples in the literature:
we have $\langle n\rangle\approx 5.1$ (with $s_n \approx 1.1$)
 compared with $\langle n\rangle \approx 3.6$ 
for 260 ATLAS$^{\rm 3D}$ galaxies (\citealt{Kraj}). (See further section~6.)
 This sample will be our standard (fiducial) choice 
because our analysis is based on the spherical Jeans equation.
Other samples will be used to estimate systematic errors of galaxy sampling.
 In particular, we also consider a $C_r > 2.6$ subsample of 2,607 
 nearly spherical and disk-less ellipticals
 to see the effects of varying the ETG selection criterion. 
The $C_r > 2.6$ subsample has a somewhat lower S\'{e}rsic mean,  
$\langle n\rangle\approx 4.7$.

The above measured parameters are uncertain to varying degrees.
Our formal estimates of the measurement errors are: 
$s_{\log \sigma_{\rm ap}} \approx 0.04$; $s_{\log M_{\star}^{\rm Ch}} \approx 0.1$,
$s_{\log M_{bulge,\star}^{\rm Ch}} \approx 0.2$; 
$s_{\log R_{\rm eff}} \approx 0.04$ at 
$\log_{10} (M_\star^{\rm Ch}/{\rm M}_\odot) =10.5$ varying to $\approx 0.07$ at
$\log_{10} (M_\star^{\rm Ch}/{\rm M}_\odot) =11.3$ 
for the galaxy (i.e.\ a single S\'{e}rsic-fit galaxy or a bulge plus a disk in 
the SerExp-fit case); but,
 $s_{\log R_{\rm bulge,eff}} \approx 0.09$ at 
$\log_{10} (M_\star^{\rm Ch}/{\rm M}_\odot) =10.5$ varying to $\approx 0.12$ at
$\log_{10} (M_\star^{\rm Ch}/{\rm M}_\odot) =11.3$
 for the bulge component in the SerExp fit. 
Since these errors are our formal estimates, we consider increasing them
by 0.05~dex.
As discussed in appendix~A of \cite{Ber} the estimate of $M_\star^{\rm Ch}$
depends on the method used. Our fiducial choice is that by \cite{Bel03} which
is similar to \cite{Gal05} but gives $\approx 0.1$~dex systematically larger 
$M_\star^{\rm Ch}$ compared with \cite{BR07}.
Furthermore, stellar IMFs are not universal as described in section~2.3. 
Hence stellar mass is relatively more uncertain than other parameters and
should be treated with caution. 
The aperture velocity dispersion $\sigma_{\rm ap}$ has the smallest error and
is the most reliable parameter out of our measurements.

\subsection{Stellar initial mass functions of ETGs}

A number of recent studies (e.g., \citealt{Cap13b,CvD,Tor}) find that the 
stellar IMFs for ETGs show significant galaxy-to-galaxy scatter and are in 
general different from the IMFs inferred for the Milky Way and other nearby 
galaxies (\citealt{Chab,Krou}). This means that the stellar masses derived for 
our galaxies using the Chabrier IMF should be corrected.
 Here we do not intend to describe recent IMF results 
comprehensively but just list some results that we use
to encompass the current likely range. 

\begin{figure} %2
\begin{center}
\setlength{\unitlength}{1cm}
\begin{picture}(11,8)(0,0)
\put(-1.,8.){\includegraphics{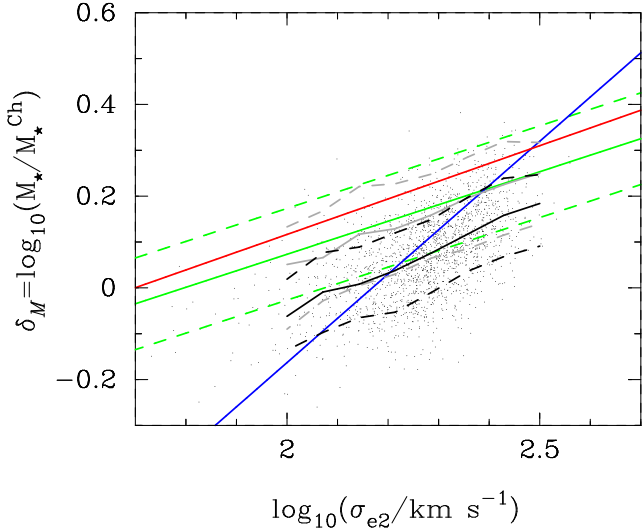}}
\end{picture} 
\caption{
Stellar IMF variation as a function of $\sigma_{\rm e2}$, aperture velocity
dispersion normalized to a radius of $R_{\rm eff}/2$.
$M_\star$ is the stellar mass of the galaxy based on the 
$\sigma_{\rm e2}$-dependent IMF
 while $M_\star^{\rm Ch}$ is that based on the Chabrier IMF. Green solid and 
dashed lines are respectively the mean relation and intrinsic scatters from the
ATLAS$^{\rm 3D}$ project. The red solid line is the linear fit relation
of the SPS modelling results of local ETGs  while the blue solid line is
an approximate median of two linear fit relations with and without modified
adiabatic contraction of DM distribution from the SPIDER project. 
Black points and lines represent the distribution from modelling $\sim$~2,000 
nearly spherical galaxies in our standard sample based on the ATLAS$^{\rm 3D}$ 
input and other standard inputs (see Table~1) while gray lines represent an 
alternative distribution through a second modelling method for the same 
inputs (see section~3.3 and section~4.2.2). Note that for the literature
 results velocity dispersions have been converted to
$\sigma_{\rm e2}$ assuming $\sigma_R \propto R^{-0.06}$. 
}
\label{IMF}
\end{center}
\end{figure}

 The IMFs of ETGs show systematic trends with galaxy parameters such as
velocity dispersion, mass-to-light ratio ($M/L$) and 
magnesium-to-iron abundance ratio [Mg/Fe]. 
We use the VD-dependent IMF derived by the ATLAS$^{\rm 3D}$ project
from detailed dynamical modelling of 260 nearby ETGs using observed light 
distributions and velocity moments (\citealt{Cap13a,Cap13b}). 
The ATLAS$^{\rm 3D}$ IMF for each galaxy corresponds to the global IMF for the 
galaxy (i.e.\ for a region of radius $\gtrsim R_{\rm eff}$).  
The ATLAS$^{\rm 3D}$ IMFs are heavier than the Chabrier and Kroupa IMFs for most
 range of VD and become heavier slowly with increasing VD (Fig.~\ref{IMF}).
An independent IMF  distribution is obtained by detailed stellar 
population synthesis (SPS) 
modelling of the spectra of 34 nearby ETGs (\citealt{CvD}). 
The SPS modelling has nothing to do 
with galaxy dynamics so that the result is independent of the unknown DM 
distribution. The SPS modelling result corresponds to the central cylindrical 
region of radius $R_{\rm eff}/8$. The SPS linear relation shown in Fig.~\ref{IMF}
 is a least-square fit result based on the data including measurement errors
  (C.~Conroy, personal communications).
This result is similar to the ATLAS$^{\rm 3D}$
 result indicating that the IMF does not significantly vary radially within the
 galaxy. IMFs varying more rapidly than these are obtained by the SPIDER 
project through a combined dynamical and SPS modelling of a large number
($\sim 4500$) of ETGs (\citealt{Tor}). 

Considering the agreement between the ATLAS$^{\rm 3D}$ and the SPS results we
assume that there is no radial gradient of IMF within a galaxy. We use the
ATLAS$^{\rm 3D}$ IMF distribution as our standard choice and use the others
 to estimate systematic errors due to IMF uncertainties.

\subsection{Stellar mass--halo mass relation for ETGs}

\begin{figure} %3
\begin{center}
\setlength{\unitlength}{1cm}
\begin{picture}(8,10)(0,0)
\put(-0.5,0.){\includegraphics{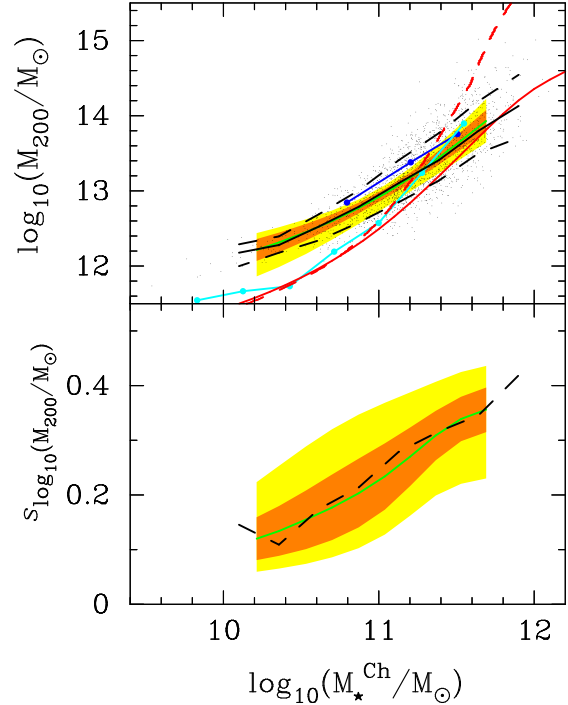}}
\end{picture} 
\caption{$M_\star^{\rm Ch}$-$M_{200}$ relation and its intrinsic 
scatter. Orange and yellow regions show respectively the 68\% and 95\% 
confidence ranges of the mean and intrinsic scatter of 
$\log_{10}(M_{200}/{\rm M}_\odot)$ as a function of $M_\star^{\rm Ch}$ 
from satellite kinematics of SDSS ETGs (More et al.\ 2011).
 Blue and cyan curves represent the mean relations
from weak lensing (Schulz et al.\ 2010; Mandelbaum et al. 2006). 
Red solid curve is the mean of $M_{200}$ at fixed $M_\star^{\rm Ch}$ 
while the red dashed curve is the mean of $M_\star^{\rm Ch}$ at fixed $M_{200}$ 
from abundance matching (Chae et al.\ 2012). Black points represent 
the distribution from modelling $\sim 2,000$ nearly spherical
galaxies in our standard sample based on the satellite kinematics input and 
other standard inputs (see Table~1).
}
\label{MsMh}
\end{center}
\end{figure}

Although there are many observational and abundance matching results for
the stellar mass ($M_\star$)--halo mass ($M_{200}$) relation for the total
population of galaxies, the results for ETGs are relatively few. We use
the $M_\star$-$M_{200}$ relations for red galaxies or ETGs obtained by the 
satellite kinematics of the SDSS galaxies (\citealt{More}), weak lensing of the
 SDSS galaxies (\citealt{Sch,Man06}) and abundance matching of the SDSS 
galaxies (\citealt{Chae12}) with the Bolshoi $N$-body simulation haloes
(\citealt{Kly}). 
The results from the satellite kinematics and 
abundance matching include intrinsic system-to-system scatter. 
These results are shown in Fig.~\ref{MsMh}. The satellite kinematics result is 
consistent with the more recent weak lensing result by \cite{Sch}, while the 
abundance matching result with the earlier weak lensing result by \cite{Man06}.
Our standard choice is the satellite kinematics result while we use the 
weak lensing and abundance matching results to estimate systematic errors. 

\subsection{Halo mass--concentration relation}

One of the current robust results from weak lensing is that the outer halo can
be well described by the NFW or Einasto profile (\citealt{Man08}). The scale 
radius or equivalently the concentration of this outer NFW/Einasto halo has 
been determined empirically by weak lensing but can be also predicted by 
cosmological $N$-body simulations.  Our standard choice is the weak lensing 
result (equation~7 and fit~2 in Table~2 of \citealt{Man08})
given by
\begin{equation}
c_{\rm NFW} (M_{200})= \frac{c_0}{1+z} \left( \frac{M_{200}}{M_0} \right)^{-\nu} 
\label{eq:McNFW} 
\end{equation}
with $c_0=5.61\pm 0.85$, $\nu=0.13\pm 0.07$ and $M_0=10^{14}h^{-1}{\rm M}_{\odot}$.
 Since some $N$-body simulation results can differ by $\sim 2\sigma$ from the 
weak lensing result, we use $\pm 2\sigma$ ($\approx \pm 30\%$)
errors of the mean concentration to estimate systematic errors.
Our adopted range is consistent with recent weak lensing measurements
(e.g.\ \citealt{Bri}) as well as recent $N$-body simulation results 
(e.g.\ \citealt{Mac,Duf,Pra}).
We assume an intrinsic halo-to-halo scatter of 0.1~dex from the mean relation
as $N$-body simulations predict similar scatter. Our results have little 
sensitivity on the precise value of the intrinsic scatter.

\subsection{Line-of-sight velocity dispersion profiles for ETGs}

The measured luminosity-weighted LOSVDs of ETGs can be, in most cases, well 
described by a simple power-law profile (see equation~\ref{eq:lwvd} below) 
in the optical region. We collect available data and do a uniform 
least-square-fit of LOSVD profiles. We consider the profile data for
48 nearby ETGs by \cite{Cap06}, the on-line data for 35 
Coma cluster ETGs by \cite{Meh} and the published data for 7 BCGs at a mean 
redshift of $\langle z \rangle \approx 0.25$ by \cite{New}.
The fitted values of $\eta$ (profile slope; equation~\ref{eq:lwvd} below)
 for all 90 ETGs are displayed in Fig.~\ref{VPdist}. 

\begin{figure} %4
\begin{center}
\setlength{\unitlength}{1cm}
\begin{picture}(10,7)(0,0)
\put(-1.,7.2){\includegraphics{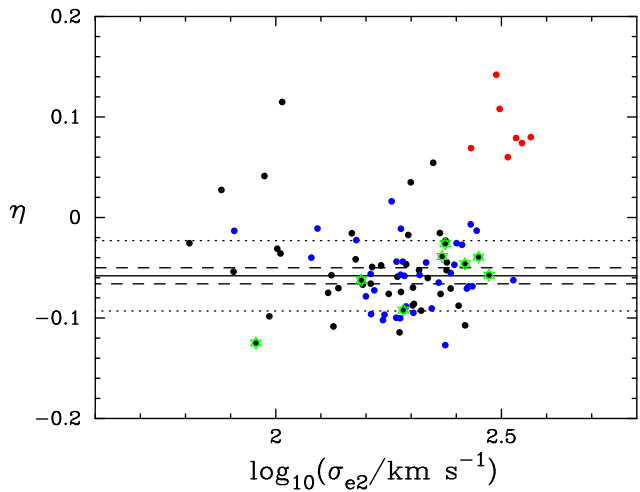}}
\end{picture}
\caption{Distribution of velocity dispersion profile slope $\eta$
(equation~11) of published nearby ETGs. 
Black points are field and cluster ETGs from Cappellari et al.\ (2006)
while blue points are ETG members of the Coma cluster (Mehlert et al.\ 2000).
 Green open stars represent nearly spherical galaxies with little rotation.
Red points are BCGs at $z \sim 0.25$ by Newman et al.\ (2013).
Black solid line is the mean of the black and blue data points with 
$\log_{10}(\sigma_{\rm e2}/{\rm km}~{\rm s}^{-1}) > 2$.
Dashed lines represent our adopted systematic errors of the mean while 
dotted lines represent intrinsic scatters of the distribution.
}
\label{VPdist}
\end{center}
\end{figure}

We note some features from Fig.~\ref{VPdist}. First, two independent samples, 
black points (\citealt{Cap06}) and blue points (\citealt{Meh}), have 
indistinguishable distributions of $\eta$. Second, low VD 
($\sigma_{\rm e2} \lesssim 100~{\rm km}~{\rm s}^{-1}$) ETGs appear to have a 
higher mean of $\eta$.  These low-$\sigma_{\rm e2}$ ETGs
tend to have relatively less dominating bulges 
(M.~Cappellari, personal communications)
meaning that they may not match well with our bulge-dominating ETGs.
 Third, all $z \sim 0.25$  BCGs have positive values of  $\eta$ 
(mean $\langle\eta\rangle_{\rm BCG} \approx 0.087$)
in contrast to Coma cluster cD and D galaxies 
NGC~4874 ($\eta=-0.0470\pm 0.0055$),  NGC~4889 ($\eta=-0.0624\pm 0.0066$)
and NGC~4839 ($\eta=-0.0068\pm 0.0134$).  The difference between the
$z \sim 0.25$ BCG sample and the Coma cD/D galaxies can be largely attributed 
to systematic size difference and redshift evolution. 
 The $z \sim 0.25$ BCGs have much larger sizes than the 
Coma cD/Ds: 5 out of the 7 BCGs have 
$30~{\rm kpc}\lesssim R_{\rm eff} \lesssim 50~{\rm kpc}$
while 2 out of the 3 Coma cD/Ds have $R_{\rm eff}< 20~{\rm kpc}$. 
Evolution effects cannot be neglected for the $z \sim 0.25$ clusters because
of the rapid redshift evolution of massive clusters.
Fourth, other than the low $\sigma_{\rm e2}$ ETG and $z \sim 0.25$ BCG 
peculiarities $\eta$ does not vary with $\sigma_{\rm e2}$. 
Finally, 8 nearly spherical 
($\epsilon <0.15$) ETGs (marked by open stars) with little rotation have a mean
 consistent with that for all ETGs. 

Excluding the  $z \sim 0.25$ BCGs, we have a mean value 
$\langle\eta\rangle=-0.0527 \pm 0.0045$ with an intrinsic scatter of
 $s_{\eta}= 0.0406$ for 83 ETGs and a slightly different value 
$\langle\eta\rangle=-0.0579 \pm 0.0039$ with $s_{\eta}= 0.0340$ for
77 ETGs with $\sigma_{\rm e2} > 100~{\rm km}~{\rm s}^{-1}$. For the 8 
nearly spherical and slowly rotating ETGs we have 
$\langle\eta\rangle=-0.061 \pm 0.012$ with $s_{\eta}= 0.031$.
These values are consistent with one another. Our fiducial choice is
 $\langle\eta\rangle=-0.058$ with $s_{\eta}= 0.035$. This value 
is somewhat higher than the value $\langle\eta\rangle=-0.066$ by
\cite{Cap06} but the intrinsic scatter is identical.
 We take the difference between the two values (i.e.\ $0.008$) as our 
estimate of the systematic error.

\subsection{Velocity dispersion anisotropies}

Detailed dynamical modelling results of nearby ETGs show that VD
 anisotropies are varying in general with $r$ in the optical 
region (e.g., \citealt{Ger,Tho,Cap07}). 
 Despite the radial variations anisotropy 
 $\beta(r)$ (see below equation~\ref{eq:jeans} and the subsequent text)  
values are bounded within the 
range $-0.9 \lesssim \beta(r) \lesssim 0.5$ for the entire radial range for 
most ETGs. For the case of using constant anisotropies we use an asymmetric
pseudo-Gaussian probability density distribution given by 
\begin{equation}
P(\beta) = \frac{1}{\sqrt{\pi/2}(s_{\rm L}+s_{\rm H})}\times
        \left\{\begin{array}{c} 
 \exp\left[-\frac{(\beta-\mu_\beta)^2}{2s_{\rm L}^2}\right]
  \hspace{1ex} (\beta<\mu_\beta) \smallskip \\ 
 \exp\left[-\frac{(\beta-\mu_\beta)^2}{2s_{\rm H}^2}\right]
  \hspace{1ex} (\beta\ge\mu_\beta)
   \end{array} \right.
\label{eq:PDFbet}
\end{equation}
with $\mu_\beta = 0.18$ (peak value), $s_{\rm H}=0.11$ (higher value dispersion) 
and $s_{\rm L}=0.25$ (lower value dispersion) derived by \cite{Chae12} using 40 
unoverlapping ETGs in the literature. The distribution has a mean of 
$\langle\beta\rangle = 0.056$. 
We take this constant value distribution as our fiducial choice.

As a way of estimating  systematic effects of radially varying anisotropies 
we use a smoothly varying function of the form 
(appendix~B, \citealt{Chae12})
\begin{equation}
\beta(r)=\beta_0+ \beta_1\frac{r^2}{r^2+r_1^2}+ \beta_2\frac{r^2}{r^2+r_2^2}
\label{eq:varbet}
\end{equation}
with the central anisotropy $-0.3< \beta_0< 0.3$, $0 <r_1, r_2 <R_{\rm eff}$, 
mean anisotropy within $R_{\rm eff}$ taking the distribution of 
equation~(\ref{eq:PDFbet}), and the anisotropy at large radii 
$0<\beta_\infty<0.3$ motivated from N-body simulations 
(e.g.\ \citealt{Nav10}).

\subsection{The centres of early-type galaxies}

While the main bodies of ETGs can be well described by the S\'{e}rsic (or 
SerExp) model, the central sub-kiloparsec regions are known to deviate from the 
the S\'{e}rsic extrapolation (e.g., \citealt{Fab,Res,Hyd,Kor,Gla}). 
There can be missing or extra
 light compared with the S\'{e}rsic extrapolation of the bulge component (or
the galaxy) depending on the nature of the galaxy. 
The galaxies with missing central light are mostly giant ETGs 
(sometimes referred to as the core galaxies), while those with extra central 
light are less luminous ETGs (sometimes referred to as the power-law galaxies).
On average, core galaxies have much shallower central slopes compared with 
power-law galaxies.

\begin{figure} %5
\begin{center}
\setlength{\unitlength}{1cm}
\begin{picture}(10,7)(0,0)
\put(-0.6,7.1){\includegraphics{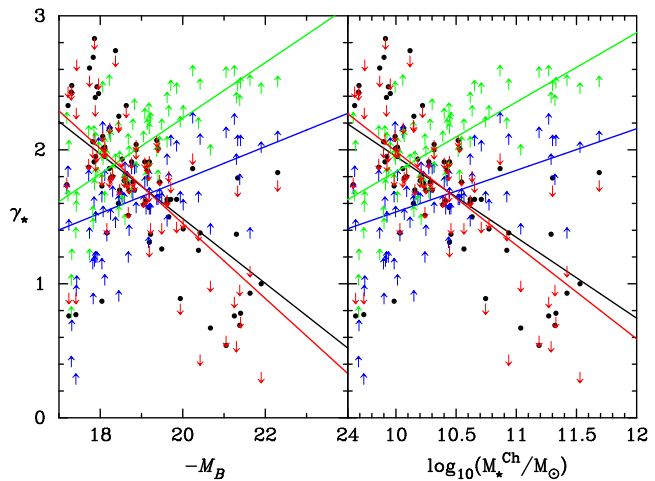}}
\end{picture} 
\caption{Measured values of the three-dimensional luminosity (stellar mass) 
density slope $\gamma_\star (\equiv - d \ln\rho_\star(r)/d\ln r)$ at
four different radii, $0.005 R_{\rm eff}$ (red symbols), 
$0.01 R_{\rm eff}$ (black), $0.05 R_{\rm eff}$ (blue) and $0.3 R_{\rm eff}$ (green)
for 79 ETGs ($M_B < -17$) imaged with the \emph{Hubble Space Telescope} 
ACS (Glass et al.\ 2011). In the right-hand side panel $M_\star^{\rm Ch}$ is 
the stellar mass based on the Chabrier IMF using a
photometric conversion described in the text.
}
\label{gamMBMs}
\end{center}
\end{figure}

Recent detailed \emph{Hubble Space Telescope} observations of the centres of 
 ETGs for a wide range of luminosities show that the central slope  varies 
systematically with luminosity (\citealt{Gla,Res}). Furthermore, the central 
slope varies radially in the central region of a galaxy. 
Fig.~\ref{gamMBMs} shows three-dimensional central slope 
$\gamma_\star$ ($\equiv - d \ln\rho_\star(r)/d\ln r$) values at four 
different radii, $0.005 R_{\rm eff}$, 
$0.01 R_{\rm eff}$, $0.05 R_{\rm eff}$ and $0.3 R_{\rm eff}$ for 79 ETGs with 
$M_B < -17$ (\citealt{Gla}), or 
$\log_{10}(M_{\star}^{\rm Ch}/{\rm M}_\odot) \gtrsim 9.6$ using 
$\log_{10}(M_{\star}^{\rm Ch}/{\rm M}_\odot)=1.097 (g-r)-0.406-0.4 (M_r -4.67)$,
 $g-r \sim 0.75$ (\citealt{Ber}) and $B=g+0.313(g-r)+0.227$ (\citealt{Lup}).
The slope at $r=0.01 R_{\rm eff}$ scales as 
$\gamma_{0.01} = 1.34 - 0.603 (\log_{10}(M_{\star}^{\rm Ch}/{\rm M}_\odot)-11)$
(black line in Fig.~\ref{gamMBMs}).

\begin{figure} %6
\begin{center}
\setlength{\unitlength}{1cm}
\begin{picture}(10,7)(0,0)
\put(-0.5,7.1){\includegraphics{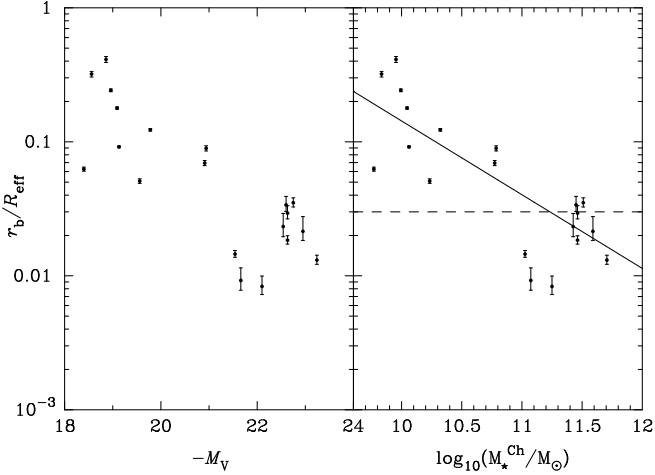}}
\end{picture} 
\caption{Measured values of break radius $r_{\rm b}$ normalized 
by $R_{\rm eff}$ for 21 ETGs of the Virgo cluster brighter than $M_V=-18$ based on
a wealth of data taken with a number of telescopes including the 
\emph{Hubble Space Telescope} (Kormendy et al.\ 2009). 
Here each value of $r_{\rm b}$ is defined to be the minimum radius of the 
main-body S\'{e}rsic fit except for NGC~4459 whose
 minimum radius was reread from its light profile.}
\label{rb}
\end{center}
\end{figure}

The break radius at which the luminosity profile starts to deviate from the 
main-body S\'{e}rsic profile is typically 
$r_{\rm b}\approx 0.03 R_{\rm eff}$ (\citealt{Fab}) 
but varies systematically from $\sim 0.01 R_{\rm eff}$ up to 
$\sim 0.3 R_{\rm eff}$ according to a more recent study by \cite{Kor}. 
Fig.~\ref{rb} shows a systematic trend of $r_{\rm b}/R_{\rm eff}$ based on 21 ETGs
 of the Virgo cluster brighter than $M_V=-18$ (\citealt{Kor}). 
Stellar mass $M_{\star}^{\rm Ch}$ for each galaxy is
obtained using a photometric conversion 
$V=g-0.5784\times (g-r)-0.0038$ (\citealt{Lup}). 
We obtain a least-square fit relation 
$\log_{10}(r_{\rm b}/R_{\rm eff})=a+b \log_{10}(M_{\star}^{\rm Ch}/{\rm M}_\odot)$ 
with $a=4.6577\pm 0.0903$ and $b=-0.5502\pm 0.0088$.

The central missing or extra light cannot affect significantly our 
modelling results for the mass profile outside the central region. Nevertheless,
 we take into account the central light distribution as follows to obtain as 
accurate as possible results particularly for the central region. We assume a 
linearly varying slope $\gamma(r)= \gamma_0 + m (r/R_{\rm eff})$ with the 
corresponding density 
$\rho_\star(r)= \rho_0 (r/R_{\rm eff})^{-\gamma_0} \exp[-m(r/R_{\rm eff})]$ 
for $r < r_b$. For each galaxy in our sample 
we determine $\gamma_0$ and $m$ by requiring that the
slope is continuous with the S\'{e}rsic slope at $r_b$ and the slope at 
$r=0.01 R_{\rm eff}$ matches the empirical mean value $\gamma_{0.01}$ given above.
Our fiducial choice for $r_b$ is $0.03 R_{\rm eff}$. To estimate systematic 
errors we consider alternatively the varying $r_b$ shown in Fig.~\ref{rb}.

\begin{table*}
\caption{Summary of the standard inputs
\label{stand}}
\begin{tabular}{ccl}
\hline 
\hline 
  item   &  description & reference \\
\hline
\hline 
galaxy sample & \( \begin{array}{c} 
2,054~{\rm galaxies~with~}\varepsilon<0.15 \\
 \& \log_{10}(R_{\rm eff}^a/R_{\rm bulge,eff}^b) < 0.19 \end{array} \)
 & this work  \\
\hline 
stellar IMF distribution & ATLAS$^{\rm 3D}$ & \cite{Cap13b}\\
\hline 
$M_\star^{\rm Ch}$-$M_{200}$ relation &  satellite kinematics  & 
\cite{More} \\
\hline 
\(\begin{array}{c}M_{200}-c_{\rm NFW}~{\rm relation}\\
 {\rm for}~r>0.2 r_{200}\end{array}\) &
 \(\begin{array}{c} {\rm weak~lensing~relation} \\
  {\rm equation}~(\ref{eq:McNFW}) \\
 {\rm with~intrinsic~scatter~of~0.1dex}
\end{array}\) & \cite{Man08} \\
\hline 
  VD profile & $\langle\eta\rangle = -0.058$, $\sigma_\eta=0.035$ & 
 derived~from~the~literature   \\ 
\hline 
 VD anisotropy  &  \(\begin{array}{c}
      {\rm asymmetric~pseudo-Gaussian~distribution} \\
      (\mu_\beta=0.18, \sigma_{\rm H}=0.11, \sigma_{\rm L}=0.25)\\   
    {\rm of~constant}~\beta~{\rm (equation~\ref{eq:PDFbet})} \end{array}\) &
   derived from the literature \\
\hline 
 break (core) radius & $r_b=0.03 R_{\rm eff}$  & \cite{Fab} \\
\hline 
 DM distribution model & gNFW   &  generic (equation~\ref{eq:rhogNFW}) \\
\hline
\hline 
\end{tabular}

$^a$ The effective radius of the total light in the single Sersic fit. \\
$^b$ The effective radius of the bulge component in the SerExp fit.

\end{table*}

\section{Mass models and Jeans analysis}

 We use the spherical Jeans equation (see section~3.2) to constrain 
galactic mass profiles of nearly spherical galaxies in our sample. 
We take parametric approach and consider for the total mass distribution 
single-component and two-component models. 
Two-component models are clearly more realistic and better-motivated 
because observed elliptical galaxies are believed to have at least two 
separate components (i.e.\ luminous and DM components). 
In single-component models the unknown DM component is treated implicitly. 
Single-component models are considered because they can provide 
qualitatively different profiles and thus can allow useful comparison with
two-component models.

\subsection{Parametric models of mass distribution}

 In two-component models the total mass distribution results from the 
superposition of the empirically derived stellar mass distribution and the 
unknown DM distribution (gas mass is negligible for our ellipticals), namely 
\begin{equation}
 \rho(r) = \rho_\star(r) + \rho_{\rm DM}(r).
\label{eq:rhotot}
\end{equation}
 We assume that the DM profile varies smoothly from the inner
 to the outer halo. Specifically, we consider 
two classes of profiles for DM distribution.

One is the generalized NFW (gNFW) model given by 
\begin{equation}
\rho_{\rm gNFW}(r)=\frac{\rho_{\rm s}} {\left(r/r_{\rm s}\right)^\alpha 
\left(1+r/r_{\rm s}\right)^{3-\alpha}},   
\label{eq:rhogNFW}
\end{equation}
where $\alpha$ is the limiting slope as $r \rightarrow 0$, and 
$\rho_{\rm s}$ and  $r_{\rm s}$  are  related to the 
concentration parameter $c \left(\equiv r_{200}/r_{\rm s}\right)$ and $M_{200}$ 
(\citealt{Chae12}). 

The other is the Einasto model (\citealt{Ein}) given by
\begin{equation}
\rho_{\rm Einasto}(r)= \rho_{-2} 
\exp\left\{-(2/\tilde{\alpha})\left[(r/r_{-2})^{\tilde{\alpha}}-1\right]\right\},  
\label{eq:rhoEin}
\end{equation}
where $r_{-2}$ (the radius at which the logarithmic slope of the density is 
$-2$) and $\rho_{-2}$ are similarly related to 
$c_{-2}\left(\equiv r_{200}/r_{-2}\right)$ and $M_{200}$ (\citealt{Chae12}).

For the single-component case we consider for the total mass distribution
the gNFW model (equation~\ref{eq:rhogNFW}) and a simple power-law model given by
\begin{equation}
\rho_{\rm PL}(r)= \rho_{0} \left( \frac{r}{r_0} \right)^{-\gamma_{\rm PL}},  
\label{eq:rhoPL}
\end{equation} 
where $\rho_{0}$ is the density at a fiducial radius $r_0$.
This constant slope model (equation~\ref{eq:rhoPL}) is not intended to describe
 mass profile at large radii but just the profile in the optical region.

\subsection{Spherical Jeans equation and velocity dispersion}

For the total mass profile $M(r)=M_{\star}(r)+M_{\rm DM}(r)$ we use
 the spherical Jeans equation (\citealt{BT}) given by 
\begin{equation}
\frac{d[\rho_{\star}(r) \sigma_{\rm r}^2(r)]}{dr} 
+ 2 \frac{\beta(r)}{r} [\rho_{\star}(r) \sigma_{\rm r}^2(r)]
= - G \frac{\rho_{\star}(r) M(r)}{r^2},
\label{eq:jeans}
\end{equation}
where $\sigma_{\rm r}(r)$ is the radial stellar velocity dispersion at radius 
$r$ and $\beta(r)$ is the velocity dispersion anisotropy at $r$ given by
$\beta(r)=1 - \sigma_{\rm t}^2(r)/\sigma_{\rm r}^2(r)$
where $\sigma_{\rm t}(r)$ is the tangential velocity 
dispersion in spherical coordinates. An integral solution of the Jeans 
equation for $\sigma_{\rm r}(r)$ with the general form of $\beta(r)$ given by 
equation~(\ref{eq:varbet}) can be found in \cite{Chae12}.

The LOSVD of stars at projected radius $R$ on the sky $\sigma_{\rm los}(R)$ is 
given by (\citealt{BM})
\begin{equation}
\sigma_{\rm los}^2(R)=\frac{1}{\Sigma_{\star}(R)} \int_{R^2}^{\infty}
\rho_{\star}(r) \sigma_{\rm r}^2(r) \left[ 1 - \frac{R^2}{r^2} \beta(r) \right]
 \frac{dr^2}{\sqrt{r^2-R^2}},
\label{eq:losvd}
\end{equation}
where $\Sigma_{\star}(R)$ is the two-dimensional stellar mass density projected
on the sky. Observed ETGs show that 
the luminosity weighted LOSVD within an aperture 
of radius $R$ is, in most cases, well-described by a 
power-law profile (\citealt{Cap06,Meh}), i.e.,
\begin{eqnarray}
\sigma_R \equiv \langle \sigma_{\rm los} \rangle (R) & = &
 \frac{\int_0^R \Sigma_{\star}(R') \sigma_{\rm los}(R') R' dR'}
{\int_0^R \Sigma_{\star}(R') R' dR'}    \nonumber  \\
 & = & \sigma_{\rm e2} \left (\frac{R}{R_{\rm eff}/2} \right)^\eta,
\label{eq:lwvd}
\end{eqnarray}
 where $\sigma_{\rm e2}$ is the velocity dispersion within the fiducial 
radius of $R_{\rm eff}/2$, which on average corresponds to the 
aperture radius for the SDSS galaxies in our sample 
(see Fig.~\ref{Rz}).

\subsection{Procedures of constructing a model set}

 For each galaxy in our sample the only direct dynamical constraint is
 the aperture velocity dispersion $\sigma_{\rm ap}$. Additionally, we have 
the statistical distribution of velocity dispersion profile slopes within 
$R_{\rm eff}$ (section~2.6) and the statistical information of the halo virial 
mass $M_{200}$ (section~2.4) and its profile at $r > 0.2 r_{200}$ (section~2.5). 
The procedure of constructing a model set for our galaxy sample is not unique 
but depends on how to treat these direct and indirect constraints.
We consider two approaches. 

In the first approach, to be called ``step-by-step approach'', we produce
a degenerate parameter set of a given mass model that are solutions of the
spherical Jeans equation for $\sigma_{\rm ap}$.  Then, we use the indirect 
constraints to break the degeneracy and test the model set statistically. 
In this approach $\sigma_{\rm ap}$, the most reliable dynamical information, is
reproduced without any bias.  

In the second approach (to be called ``chi-square approach''), we try to fit 
simultaneously all direct and indirect constraints by defining a 
goodness-of-fit $\chi^2$ taking into account all available measurement errors 
and intrinsic scatters. In this approach all direct and indirect dynamical 
constraints are treated equally. However, it does not guarantee that 
 $\sigma_{\rm ap}$ is reproduced without bias.
For this reason our standard approach will be the step-by-step approach.
We describe in detail the two procedures in turn.

\subsubsection{Step-by-step approach}

\begin{enumerate}

\item Each galaxy  has the stellar mass $M_\star^{\rm Ch}$ based
 on the Chabrier IMF and S\'{e}rsic parameters fitted to the surface brightness
 data. A three-dimensional stellar mass profile is obtained by de-projecting 
the S\'{e}rsic surface brightness profile assuming the constant Chabrier IMF.

\item  (a) We assign $M_\star$ by drawing a value of 
$\delta_M \equiv \log_{10}(M_\star/M_\star^{\rm Ch})$ randomly from an empirical 
distribution of IMF as a function of velocity dispersion as described in 
section~2.3.
(b) We assign a halo mass $M_{200}$ using an empirical
stellar-to-halo mass relation (section~2.4) and an NFW concentration 
$c_{\rm NFW}$ for $r > 0.2 r_{200}$ using weak lensing derived 
$c_{\rm NFW}$-$M_{200}$ relation (section~2.5).  
(c) We also assign a VD anisotropy $\beta(r)$ 
from an empirical distribution (section~2.7).

\item For the above assignment of $\delta_M$, $M_{200}$, 
$c_{\rm NFW}$ and $\beta(r)$ to the galaxy
 we produce a degenerate parameter set $\{\alpha,c\}$ with 
the prior $0< \alpha < 3$ for the gNFW 
(equation~\ref{eq:rhogNFW}) or $\{\tilde{\alpha},c_{-2}\}$
with the prior $0< \tilde{\alpha} < 0.9$ for the Einasto
(equation~\ref{eq:rhoEin}) by requiring that each set reproduces 
$\sigma_{\rm ap}$. In some cases the assigned random parameter combinations
from step~(ii) may not allow any solution for $\sigma_{\rm ap}$. 
In those cases we go back to step~(ii) and re-assign parameters.
For about 10 percent of galaxies solutions are not found through a significant
number of iterations. We reject those galaxies in this approach.
In this random process the posterior distribution of any parameter may
 be readjusted from the prior input distribution as required by the aperture 
velocity dispersion.  We allow this readjustment to happen because the
aperture velocity dispersion is more reliable than those empirical inputs
used in step~(ii).  For the case of the power-law model
(equation~\ref{eq:rhoPL}) for the total mass distribution we have two
free parameters $\gamma_{\rm PL}$ and $\rho_0$ but we fix $\rho_0$ by
requiring that near the galactic centre the total mass density is
equivalent to the stellar mass density to a good approximation. So, in this
case the only free parameter $\gamma_{\rm PL}$ is uniquely determined by 
$\sigma_{\rm ap}$. 

\item Out of the degenerate set from step~(iii) we select a model that best 
matches the weak lensing constraint by minimizing the following 
\begin{equation}
\Delta^2 = \sum_i \left( \frac{\log\rho^{\rm mod}(r_i) -\log\rho^{\rm WL}(r_i)}
               {s_{\log\rho(r_i)}} \right)^2,
\label{eq:delsqWL}
\end{equation}
where $0.2 r_{200}< r_i < r_{200}$, $\rho^{\rm mod}(r_i)$ is the model density, 
$\rho^{\rm WL}(r_i)$ is the density from the weak lensing NFW-fit concentration
(section~2.5) for the given halo mass, and $s_{\log\rho(r_i)}$ is the
error associated with the scatter of 0.1~dex in $c_{\rm NFW}$. 

\item Finally, we calculate the slope of $\sigma_R$ profile, $\eta$ 
(equation~\ref{eq:lwvd}), for the selected model from step~(iv) using
a least-square fit between $0.1 R_{\rm eff}$ and $R_{\rm eff}$. We then compare
the distribution of $\eta$ for the galaxy sample with the observed distribution 
(Fig.~\ref{VPdist}, section~2.6) to test the mass model being considered.

\end{enumerate}

The above procedures (iii) and (iv) can be simplified by noting that
the minimization of equation~(\ref{eq:delsqWL}) is, to a good approximation, 
equivalent to a constraint relation between $\alpha$ and $c$ (or, 
between $\tilde{\alpha}$ and $c_{-2}$) for the assigned value of
$c_{\rm NFW}$ for $r> 0.2 r_{200}$.
Such a constraint relation can be obtained in 
several ways but most easily by requiring that the slope of the DM profile 
matches that of the NFW profile at an optimal radius $r_0$.

The constraint relation for the gNFW profile is 
\begin{equation}
 c=\max\left[ \left(\frac{3-\alpha}{2} c_{\rm NFW} + 
   \frac{1-\alpha}{2}\frac{r_{200}}{r_0}\right),\delta \right],
\label{eq:congNFW}
\end{equation}
where $\delta$ is an arbitrarily small positive number ensuring that $c$ is
positive, and that for the Einasto profile is
\begin{equation}
 c_{-2}=  \frac{r_{200}}{r_0} \left[ \frac{3}{2} -  
   \frac{1}{1 + c_{\rm NFW} r_0/r_{200}} \right]^{1/\tilde{\alpha}}. 
\label{eq:conEin}
\end{equation}
From numerical experiments we select $r_0 = (2/9)r_{200}$.

\begin{figure} %7
\begin{center}
\setlength{\unitlength}{1cm}
\begin{picture}(6,9)(0,0)
\put(-1.,-0.5){\includegraphics{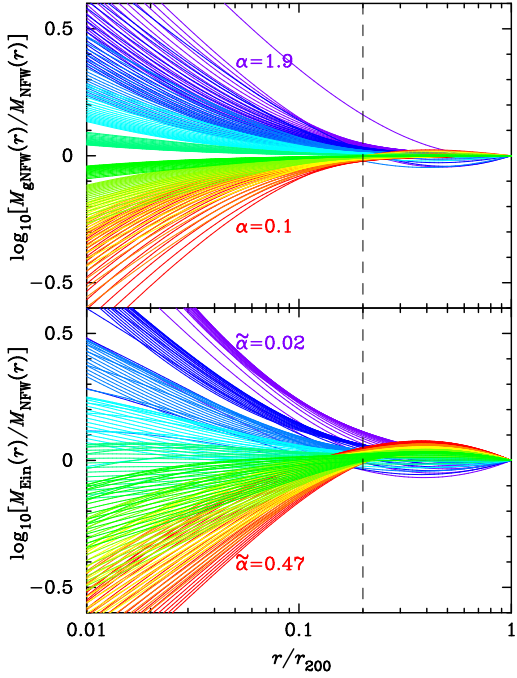}}
\end{picture}
\caption{The DM mass profiles relative to the NFW profile with
the outer halo ($r>0.2 r_{200}$) profile constraints 
(equation~14 \& equation~15) imposed. 
Each coloured set of curves are for a specific value of $\alpha$ for the gNFW or
 $\tilde{\alpha}$ for the Einasto. Different curves in the set correspond to 
different values of $c_{\rm NFW}$ from the range $2 \le c_{\rm NFW} \le 29$.
In the outer halo region the deviation of the gNFW mass $M_{\rm gNFW}(r)$ from 
the NFW mass $M_{\rm NFW}(r)$ is less than 0.05~dex for almost all occurrences 
and less than 0.02~dex for most likely occurrences. The deviation of the 
Einasto mass $M_{\rm Ein}(r)$ can be up to 0.1~dex but is less than 0.05~dex for
most likely occurrences.
}
\label{DMtoNFW}
\end{center}
\end{figure}

Fig.~\ref{DMtoNFW} shows the gNFW/Einasto mass profiles 
relative to the NFW profile with the outer profile 
constraints (equation~\ref{eq:congNFW} \& equation~\ref{eq:conEin}) imposed. 
We consider $2 \le c_{\rm NFW} \le 29$
to fully cover the empirical range. We then consider the
plausible posterior range  $0.1\le \alpha \le 1.9$ for the gNFW and 
$0.02\le\tilde{\alpha}\le 0.47$ for the Einasto. For virtually all plausible 
combinations of $c_{\rm NFW}$ and $\alpha$ the gNFW profile deviates from the
NFW by less than 0.05~dex for $0.2 r_{200}\le r \le r_{200}$. For the Einasto
profile the deviation can be as large as $\approx 0.1$~dex but for the most
likely combinations it is less than 0.05~dex.

With the constraint relation (equation~\ref{eq:congNFW} or \ref{eq:conEin})  
 the effective number of free parameters is just one. 
We choose one free parameter to be $\alpha$ (or, $\tilde{\alpha}$).
Notice that by varying the free parameter of the assumed DM
model baryonic effects in the inner halo (\citealt{Gne,SM,Gne11}), 
if present, can be taken into account.

\subsubsection{chi-square approach}

In this approach we intend to fit all three dynamical constraints (i.e.\
$\sigma_{\rm ap}$, the statistical distribution of $\eta$, and
the statistical distribution of $c_{\rm NFW}$ for $r > 0.2 r_{200}$) 
simultaneously by defining a suitable $\chi^2$. 
The power-law model (equation~\ref{eq:rhoPL}) is not applicable for this 
approach because it cannot simultaneously satisfy both stellar kinematics and 
weak lensing constraints. 
We first note that we can use the constraint relation 
(equation~\ref{eq:congNFW} or \ref{eq:conEin}) to reduce one term in $\chi^2$.
We have verified that including a term like equation~(\ref{eq:delsqWL}) is 
equivalent to using the constraint. Hence there remains just one free
parameter $\alpha$ (or $\tilde{\alpha}$) and  
we define the following goodness-of-fit statistic
\begin{equation}
\chi^2 = \chi^2_{V} + \chi^2_{\eta} 
\label{eq:chisq}
\end{equation}
where $\chi^2_{x}=\left(x^{\rm mod}-x^{\rm emp}\right)^2/s_x^2$ for
$V=\log_{10}(\sigma/{\rm km}~{\rm s}^{-1})$ and $\eta$ (equation~\ref{eq:lwvd}) 
with $x^{\rm mod}$ and $x^{\rm emp}$ are  the theoretical model and 
empirical (observational) values and $s_x$ is the error contributed by 
 the measurement error for  $x^{\rm emp}$ and/or the error for $x^{\rm mod}$ due 
to the measurement errors of $M_{\star}$ and $R_{\rm eff}$ 
 as described in section~2.2.

The first term in equation~(\ref{eq:chisq}) is 
\begin{equation}
\chi^2_{V}=\frac{\left[V^{\rm mod}(\alpha;\vec{p})-V^{\rm emp}\right]^2}
        {s_V^2},
\label{eq:chisqVD}
\end{equation}
where $V^{\rm emp}$ is the measured value and $\vec{p}$ includes the empirical 
input parameters.  The dispersion $s_V^2$ is
\begin{equation}
s_V^2 = s_{V,{\rm meas}}^2 + \delta V^2 (s_{\log R_{\rm eff}})
 + \delta V^2 (s_{\log M_{\star}}) 
\label{eq:errVD}
\end{equation}
where  $s_{V,{\rm meas}}$ is the measurement error and
$\delta V (s_{\log R_{\rm eff}})$ and $\delta V (s_{\log M_{\star}})$
 are the errors in $V^{\rm mod}$ respectively due to
$s_{\log R_{\rm eff}}$ and $s_{\log M_{\star}}$ (the measurement errors of 
$\log R_{\rm eff}$ and $\log M_\star$). 

The second term in equation~(\ref{eq:chisq}) is 
\begin{equation}
\chi^2_{\eta}=\frac{\left[\eta^{\rm mod}(\alpha;\vec{p})-
     \eta^{\rm emp}\right]^2}  {s_\eta^2}, 
\label{eq:chisqeta}
\end{equation}
where $\eta^{\rm emp}$ is a value assigned empirically (section~2.6). 
 The dispersion $s_\eta^2$ is
\begin{equation}
s_\eta^2 =  \delta \eta^2 (s_{\log R_{\rm eff}})
 + \delta \eta^2 (s_{\log M_{\star}})  
\label{eq:dispeta}
\end{equation}
where  $\delta\eta(s_{\log R_{\rm eff}})$ and $\delta\eta(s_{\log M_{\star}})$
 are the errors in $\eta^{\rm mod}$ respectively due to
$s_{\log R_{\rm eff}}$ and $s_{\log M_{\star}}$ (the measurement errors of 
$\log R_{\rm eff}$ and $\log M_\star$).

We minimize the $\chi^2$ over $\alpha$ coupled with $c$ (or, 
$\tilde{\alpha}$ coupled with $c_{-2}$) with the assigned values of the
empirical parameters as in step~(ii) of section~3.2.1.
In minimizing the $\chi^2$
we impose the following prior constraints $0 < \alpha < 3$ 
(or, $0 <\tilde{\alpha} < 0.9$). For most cases,
the $\chi^2$ has a well-defined minimum within these prior constraints for the
assigned values of the input parameters. If the minimum value of $\chi^2$ has
$\chi^2_{\rm min} > 9$ or the minimum point tries to cross the boundary of the 
prior constraint, we retry by selecting other random values of the input 
parameters from the empirical ranges.  

\section{Results}

\subsection{The case of pure stellar mass distribution based on constant IMF}

\begin{figure} %8
\begin{center}
\setlength{\unitlength}{1cm}
\begin{picture}(9,6)(0,0)
\put(-0.4,7.2){\includegraphics{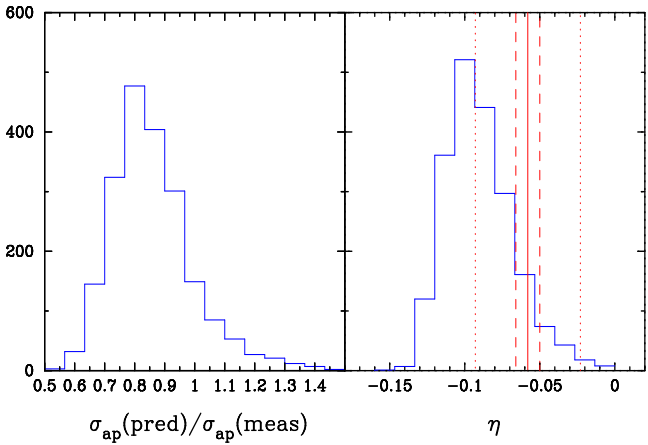}}
\end{picture}
\caption{
The predicted aperture velocity dispersion $\sigma_{\rm ap}$ and velocity 
dispersion profile slope $\eta$ (equation~12) for the pure stellar 
mass profiles of 2,054 nearly spherical galaxies in our standard sample 
based on the constant Chabrier IMF. The predicted $\sigma_{\rm ap}$
is systematically lower than the measured value while the predicted $\eta$ 
is systematically steeper than the measured distribution indicated by vertical
red lines as shown in Fig.~4.
}
\label{nomod}
\end{center}
\end{figure}

We first consider the case without any modelling.
 In this case the total mass distribution is just the
stellar mass distribution based on the constant Chabrier IMF. 
Fig.~\ref{nomod} shows the predicted aperture velocity dispersion
$\sigma_{\rm ap}$ and the velocity dispersion profile slope $\eta$ for 
$0.1 R_{\rm eff}< R < R_{\rm eff}$ based on the distribution of constant 
anisotropies given by equation~(\ref{eq:PDFbet}). We notice that the predicted 
$\sigma_{\rm ap}$ is lower and the predicted $\eta$ is steeper than measured. 
The lower $\sigma_{\rm ap}$ can, in principle, be remedied using heavier IMFs 
than the Chabrier, although this statement is uncertain due to the large 
systematic error in estimates of stellar mass even for a fixed IMF 
(see section~2.2).
However, the steeper $\eta$ can only be remedied by including another mass 
component whose relative contribution increases with radius. 
The implied mass component is of course dark matter. 

\subsection{Jeans modelling results}

Each galaxy has not only directly measured parameters such as $M_\star^{\rm Ch}$,
$R_{\rm eff}$, $n$ and $\sigma_{\rm ap}$ but also indirectly assigned parameters 
such as $\delta_M$, $M_{200}$, $c_{\rm NFW}$ and additionally $\eta$ for the
chi-square approach.
The fitted mass profile for each galaxy depends then on
 the specific set of values of the indirectly assigned parameters.
The specific set can be regarded as a member of the empirical ensemble
or multi-dimensional space of the parameters. For each galaxy there exists 
an ensemble of models corresponding to the empirical parameter ensemble. 
This means that a run of modelling for the given sample of $N$ galaxies
produces a set of $N$ models, each of which is a random member of the 
respective ensemble of models. Another run produces different members for the 
respective galaxies but the statistical properties of the sample remain the 
same.  In this sense only the statistical results of the sample are 
meaningful from our modelling. However, variations of the empirical inputs can 
change the $N$ model ensembles and accordingly the statistical results as 
discussed in section~5 below. 

In this section we present our results based on 
 the standard (fiducial) inputs summarized in Table~1. The results based on 
 other systematically varied inputs will be considered in section~5.
We first present the results for the single-component models (section~4.2.1) 
and then for the two-component models (section~4.2.2).

\subsubsection{Results for the single-component models}

For the single-component case we consider only the step-by-step approach because
the dispersion $s_\eta^2$ (equation~\ref{eq:dispeta}) is not well-defined.
Since there will be no bias in the predicted $\sigma_{\rm ap}$ by the
step-by-step approach, our primary test will be the distribution
of $\eta$. We are considering two models for the total mass distribution, 
the gNFW model (equation~\ref{eq:rhogNFW}) and the power-law model 
(equation~\ref{eq:rhoPL}). 

Due to the lack of flexibility of the single-component model 
the predicted total mass density can be lower than
 the stellar mass density near the galactic center. 
This problem cannot be avoided particularly for the gNFW model.
The constrained gNFW mass density becomes lower than the input stellar mass 
density at $R \lesssim R_{\rm eff}/3$. 
We can remedy this by considering a two-component
model explicitly including the stellar mass profile as will be done in 
section~4.2.2. 

The power-law model is intended
only as a local model for the optical region because its constant slope
cannot simultaneously describe the optical region and the halo at large radii.
So, in this case we use one remaining freedom to fix the total density near the
 centre to match the input stellar density. 

\begin{figure} %9
\begin{center}
\setlength{\unitlength}{1cm}
\begin{picture}(9,6)(0,0)
\put(-0.4,7.2){\includegraphics{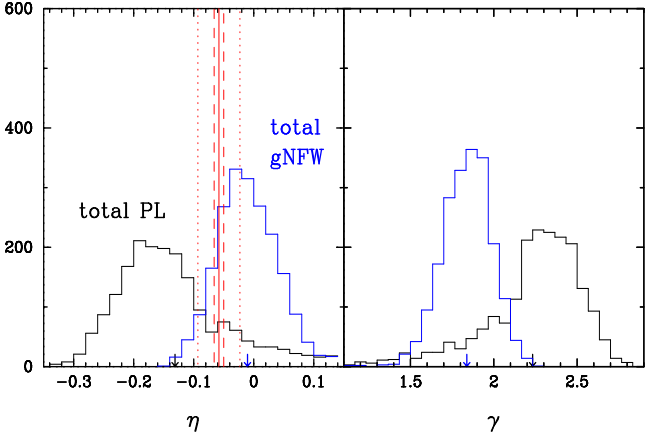}}
\end{picture}
\caption{
The predicted velocity dispersion profile slope $\eta$ (equation~12)
 and density slope $\gamma$ between $0.1 R_{\rm eff}$ and $R_{\rm eff}$ for the
single-component PL (equation~9) and 
gNFW (equation~7) models for the total mass distribution. 
The measured distribution of $\eta$ is indicated by vertical
red lines as shown in Fig.~4. The arrows indicate the mean values.
}
\label{singlemod}
\end{center}
\end{figure}

The predicted distributions of $\eta$ and $\gamma\equiv -\ln\rho(r)/d\ln r$
can be found in Fig.~\ref{singlemod}. 
These slopes are the least-square fit values between $0.1 R_{\rm eff}$ and
$R_{\rm eff}$. The predicted $\eta$ distributions are clearly discrepant with
the observed distribution and consequently the predicted $\gamma$ distributions
 are not reliable. The gNFW model predicts too shallow $\eta$ implying that
 the monotonic radial density slope variation in the model produces an artefact
  when the model is constrained by the aperture velocity dispersion at small
radii and by the weak lensing constraint at large radii.
The large scatter in $\eta$ in addition to the biased mean for the power-law 
model indicates that the model without a proper profile at large radii is
ill constrained by the aperture velocity dispersion alone. 
We will see in section~4.2.2 that the problems of the
single-component models are naturally removed when we consider more realistic
two-component models. The single-component models will not be considered any
further.

\subsubsection{Results for the two-component models}

\begin{figure} %10
\begin{center}
\setlength{\unitlength}{1cm}
\begin{picture}(9,6)(0,0)
\put(-0.4,7.2){\includegraphics{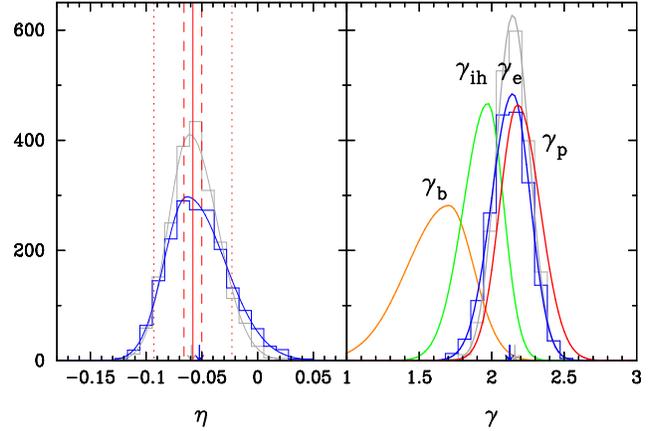}}
\end{picture}
\caption{
(Left) The predicted velocity dispersion profile slope $\eta$ 
(equation~12) between $0.1 R_{\rm eff}$ and $R_{\rm eff}$ for the
two-component model with the gNFW (equation~7) DM component with the
standard inputs (Table~1). The blue result is by the step-by-step approach 
while the gray result is by the chi-square approach. The measured distribution 
of $\eta$ is indicated by vertical red lines as shown in Fig.~4.
(Right) The predicted density slopes for the model shown left.
 $\gamma_{\rm e}$ is the least-square fit slope between $0.1 R_{\rm eff}$ and 
$R_{\rm eff}$ while $\gamma_{\rm ih}$ is between $0.1 R_{\rm eff}$ and 
$0.1 r_{200}$. $\gamma_{\rm p}$ and $\gamma_{\rm b}$ correspond 
 respectively to the peak turning point and the bottom (trough)
 turning point in a global S-shape of the slope function. 
}
\label{histetagam}
\end{center}
\end{figure}

Fig.~\ref{histetagam} shows the predicted distribution of $\eta$ for the
two-component model based on the standard inputs (Table~1). 
Its mean, median and RMS scatter are $-0.052$, $-0.054$ and $0.028$ by the
step-by-step approach and $-0.059$, $-0.060$ and $0.022$ by the chi-square
approach. The mean and median values agree well with
the observationally inferred mean $\langle\eta\rangle=-0.058\pm 0.008$. 
The predicted scatter is somewhat smaller than the observed 
scatter $s_\eta=0.035$. We show in section~5 that we can produce the
observed scatter or even larger scatters by including galaxies of  larger 
ellipticities or lower concentrations or by considering radially varying 
anisotropies. For the case of the chi-square approach 
the predicted aperture velocity dispersion is on average 0.05~dex 
higher than measured (larger than the measurement error of 0.04~dex)
 while there is no such bias for the step-by-step approach.

Fig.~\ref{IMF} shows the predicted distribution of IMF variation $\delta_M$ 
[$=\log_{10}(M_\star/M_\star^{\rm Ch})$]. The predicted slope of the variation is 
within the current empirical range. The magnitude of $\delta_M$ by the 
step-by-step approach is lower than the input ATLAS$^{\rm 3D}$ value 
(\citealt{Cap13b}) but is overall consistent with the SPIDER (\citealt{Tor}) 
value. Note that the magnitude of $\delta_M$ depends also on $M_\star^{\rm Ch}$ 
which suffers from its own uncertainty (see section~2.2). This means that the 
shift of $\delta_M$ from the input is not likely to be a problem.
The magnitude of $\delta_M$ by the chi-square approach agrees well with
the input ATLAS$^{\rm 3D}$ IMF for the sake of the biased prediction on
$\sigma_{\rm ap}$ as mentioned above.

Fig.~\ref{MsMh} shows that the posterior distribution of $M_{200}$
as a function of $M_\star^{\rm Ch}$ is consistent 
with the input relation and scatter.

\begin{figure} %11
\begin{center}
\setlength{\unitlength}{1cm}
\begin{picture}(9,7)(0,0)
\put(-0.5,8.0){\includegraphics{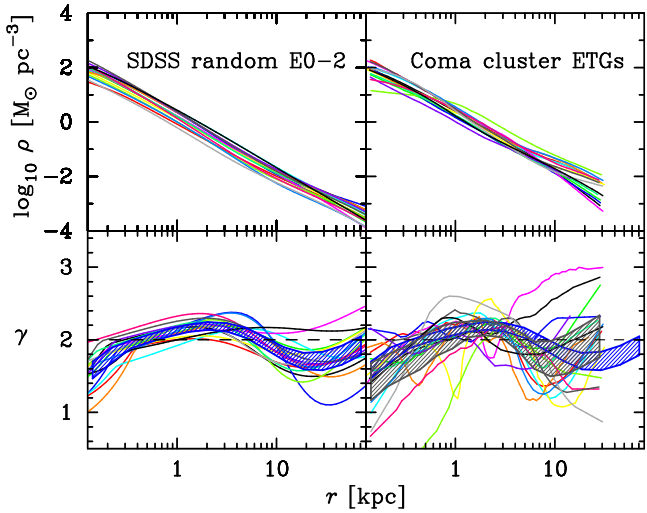}}
\end{picture}
\caption{
The left-hand side displays the radial total (stellar plus dark) mass density 
profiles for randomly selected 16 SDSS nearly spherical galaxies 
($b/a > 0.85$) and their negative slope profiles $\gamma=-d\log\rho/d\log r$. 
The blue hatched region shows the $2\sigma$ 
uncertainty of the sample mean. On the right-hand side the published  density 
profiles (Thomas et al.\ 2007) for 16 Coma cluster ETGs 
(with profiles extending beyond $R_{\rm eff}$) are compared with our ellipticals.
 The dark  hatched region shows the $2\sigma$ uncertainty of the sample mean 
for Coma cluster ETGs.
}
\label{rhogamex}
\end{center}
\end{figure}

\begin{figure*} %12
\begin{center}
\setlength{\unitlength}{1cm}
\begin{picture}(17,17)(0,0)
\put(0.5,-4.){\includegraphics{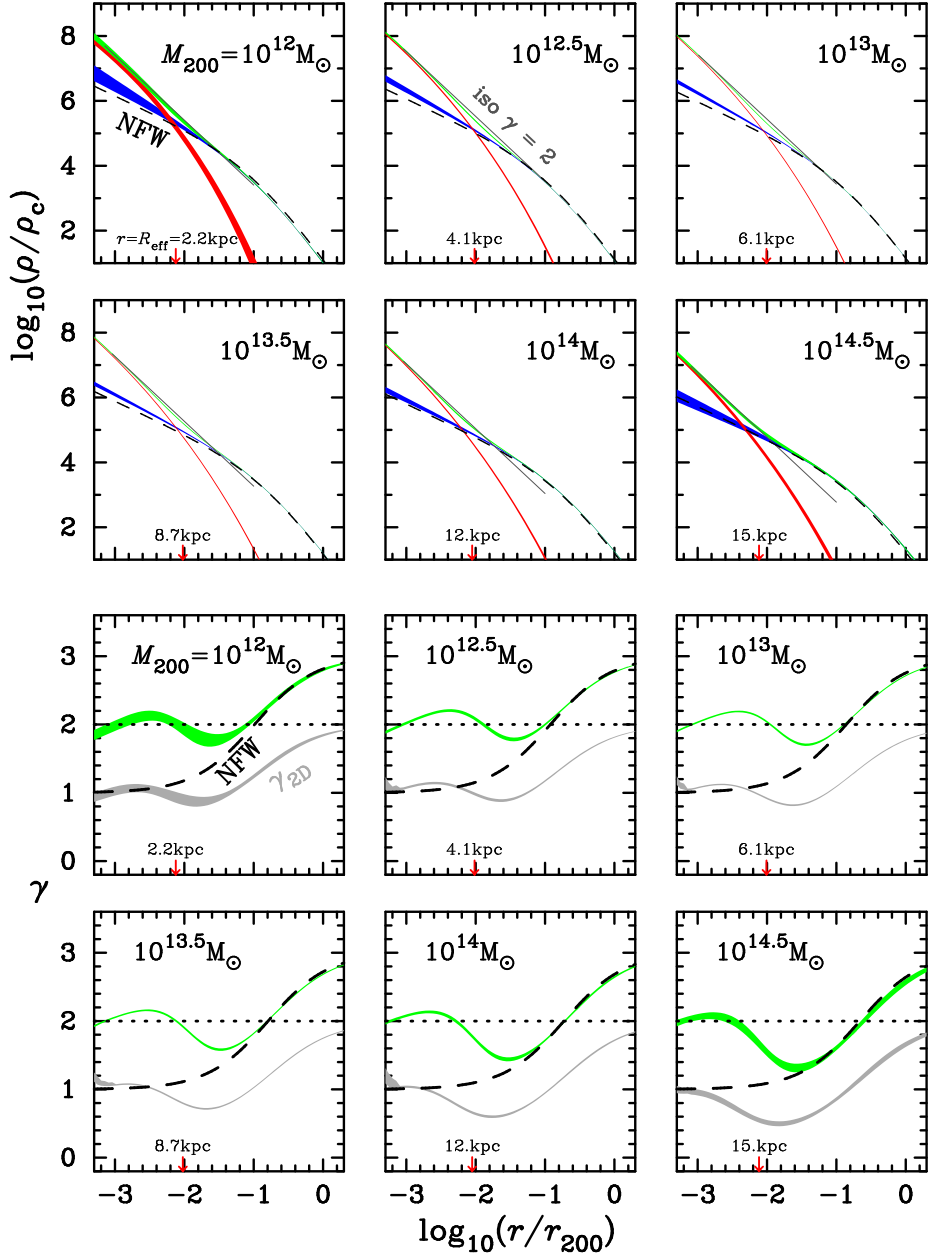}}
\end{picture}
\caption{
Density and slope profiles as in Fig.~11 for 6 sub-samples defined
by halo virial mass $M_{200}$. Each sub-sample contains a number of nearly
spherical galaxies within a range of $\pm 0.2$~dex in $M_{200}$.
 In the upper half panels, red, blue and green regions represent 
respectively the $2\sigma$ uncertainties of the sample mean density 
profiles of stellar, DM and total mass in units of 
$\rho_{\rm c}$, the critical density at $z=0.12$.
In the lower half panels, the green and gray regions represent respectively
the slope profiles for the 3-dimensional and 2-dimensional projected total 
mass densities.}
\label{rhogam}
\end{center}
\end{figure*}

For the result based on the step-by-step approach
Fig.~\ref{histetagam} shows the predicted distributions of variously
defined density slopes  $\gamma_{\rm e}$, $\gamma_{\rm ih}$,  $\gamma_{\rm p}$ 
and $\gamma_{\rm b}$, which are considered for curvatures in the 
predicted slope function.
$\gamma_{\rm e}$  is the least-square fit slope in the optical region 
between $0.1 R_{\rm eff}$ and $R_{\rm eff}$ while $\gamma_{\rm ih}$ is 
for an inner halo region between $0.1 R_{\rm eff}$ and $0.1 r_{200}$.
As shown in Fig.~\ref{rhogamex} and Fig.~\ref{rhogam} and described below, 
the predicted slope function is usually curved to have two stationary
(turning) points. $\gamma_{\rm p}$ and $\gamma_{\rm b}$ correspond to the 
values respectively at the peak and the bottom (trough) turning points.
The predicted mean and RMS scatter of $\gamma_{\rm e}$ are 
$\langle\gamma_{\rm e}\rangle \approx 2.12$ and $s_{\gamma_{\rm e}} \approx 0.14$
($\langle\gamma_{\rm e}\rangle \approx 2.16$ and $s_{\gamma_{\rm e}} \approx 0.11$
for the chi-square approach), which are steeper than isothermal.

Fig.~\ref{rhogamex} shows the total mass density profile $\rho(r)$ and 
the density slope profile $\gamma(r)$ for  16 galaxies 
 randomly selected from our sample in comparison to the published 
 density profiles of 16 Coma cluster ETGs based on detailed dynamical
modelling by \cite{Tho} (see also Fig.~14 of \citealt{Tho11}). 
There is a reasonable agreement (within $2\sigma$) between the two samples.
 There is a significant galaxy-to-galaxy scatter in the individual profiles 
but a systematic pattern in $\gamma(r)$ emerges independently from both samples.
 Typically, there is a $\gamma > 2$ (steeper than isothermal) region
within the optical region at $r < 10~{\rm kpc}$, surrounded by a $\gamma <2$ 
(shallower than isothermal) region within the inner halo ($r<0.1 r_{\rm vir}$),
which is then surrounded by a $\gamma > 2$ outer halo.

Fig.~\ref{rhogam} shows the average profiles $\rho(r)$ and $\gamma(r)$ for 6 
sub-samples defined by halo mass $M_{200}$. Each subsample has a range
of $\pm 0.2$~dex in $\log_{10} M_{200}$. This figure reveals more precisely
and in greater detail the behaviours of $\rho(r)$ and $\gamma(r)$ already 
indicated in Fig.~\ref{rhogamex}.   The average profile $\rho(r)$ in the 
optical region ($r< R_{\rm eff}$) is steeper than isothermal 
 with $\gamma \approx 2.1-2.2$ for most sub-samples. 
The optical region has a super-isothermal peak of $\gamma$ which is 
surrounded by a sub-isothermal valley in the inner halo 
($r<0.1 r_{200}$). This valley is due to the increased contribution from dark 
matter at larger radii. 
The isothermal crossing from the super-isothermal peak to the sub-isothermal
valley occurs at $r_{\rm pb}\approx R_{\rm eff}$ 
for $M_{200}=10^{13.5}{\rm M}_{\odot}$,
  $R_{\rm eff} < r_{\rm pb} < 2 R_{\rm eff}$ for $M_{200}<10^{13.5}{\rm M}_{\odot}$, and
 $0.5 R_{\rm eff}< r_{\rm pb} < R_{\rm eff}$ for $M_{200} > 10^{13.5}{\rm M}_{\odot}$.
Interestingly, if a constant slope profile
were used to fit the entire inner halo region ($0<r<0.1 r_{200}$), 
the best-fit profile would be close to isothermal for 
$M_{200} \lesssim 10^{13.5}{\rm M}_{\odot}$. Hence the isothermal model can be
 valid only as a first order approximation to the entire inner halo region
of galaxies but 
generally is not an accurate representation of the real profile.

\begin{figure*} %13
\begin{center}
\setlength{\unitlength}{1cm}
\begin{picture}(16,12)(0,0)
\put(0.,12.){\includegraphics{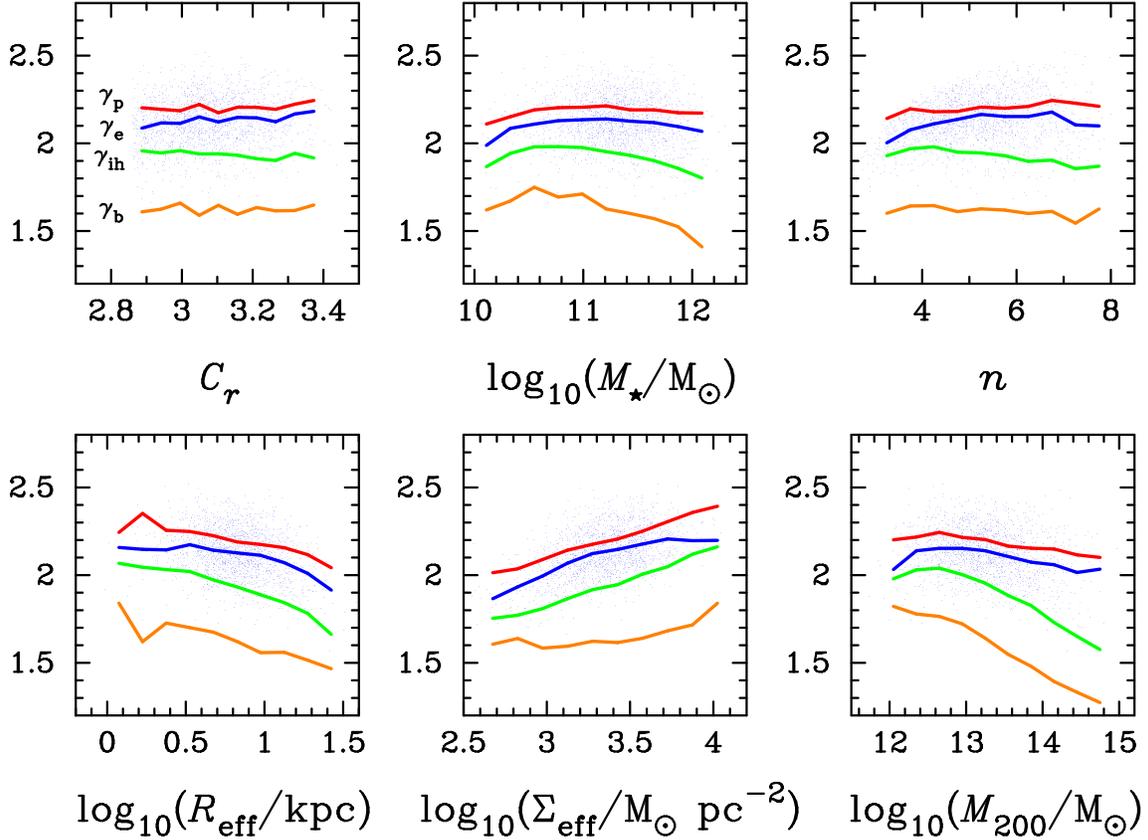}}
\end{picture}
\caption{
Four characteristic density slopes, $\gamma_{\rm p}$, $\gamma_{\rm e}$,  
$\gamma_{\rm ih}$ and $\gamma_{\rm b}$ shown in Fig.~10, against
 various parameters. Here $C_r$ is the Petrosian light concentration in 
the $r$-band as described in section~2.2 and $\Sigma_{\rm eff}$ is the projected
 stellar mass density within $R_{\rm eff}$. 
}
\label{gamvar}
\end{center}
\end{figure*}

Fig.~\ref{gamvar} shows the distribution of four characteristic slopes against
 various parameters. As indicated by the points for $\gamma_{\rm e}$, there is
significant intrinsic scatter. Other than a slow monotonic increase of 
$\gamma_{\rm e}$ with $C_r$,
the slopes have little or weak dependence on $C_r$, $M_\star$ or $n$ but 
systematically vary with $R_{\rm eff}$. In other words, galaxy size appears to be
a key indicator of global mass profiles. At a given mass, 
the smaller galaxy has a higher central density and thus a more steeply
declining profile. We also find that slopes
$\gamma_{\rm p}$, $\gamma_{\rm e}$ and $\gamma_{\rm ih}$ are well correlated
 with the projected surface density 
$\Sigma_{\rm eff}\equiv (M_\star/2)/(2\pi R_{\rm eff}^2)$ while $\gamma_{\rm b}$ is
weakly correlated at best.  The good correlation of $\gamma_{\rm e}$ with 
$\Sigma_{\rm eff}$,
which has already been noticed in the literature (e.g.\ \citealt{Aug,DT13}),
is consistent with the anti-correlation with $R_{\rm eff}$. 
The lower right panel of Fig.~\ref{gamvar} shows the distribution of the slopes
 against the host halo virial mass $M_{200}$. Slopes $\gamma_{\rm p}$ and 
$\gamma_{\rm e}$ are not well correlated with $M_{200}$. However, 
$\gamma_{\rm ih}$ declines systematically with $M_{200}$ for 
$M_{200} \ga 10^{12.5}{\rm M}_{\odot}$ and $\gamma_{\rm b}$ has a good 
anti-correlation with $M_{200}$. These behaviours of $\gamma_{\rm ih}$ and 
$\gamma_{\rm b}$ imply that central galaxies contribute less to the total mass
profile of more massive haloes. This can be attributed to the empirical
property that the stellar-to-halo mass ratio $M_\star/M_{200}$ decreases with 
increasing $M_{200}$ for $M_{200}\gtrsim 10^{12}{\rm M}_{\odot}$ 
(e.g.\ \citealt{More,Sch,Man06}) while the size-to-virial
radius ratio $R_{\rm eff}/r_{200}$ remains nearly constant (\citealt{Kra}). 
This can also be seen by Fig.~\ref{rhogam} which
shows that for more massive haloes the central stellar-to-DM density ratio 
is less boosted and consequently the NFW profile 
is retained over larger radial ranges.

\section{Systematic variations}

 We consider possible systematic variations of our statistical results 
on mass profiles
 by varying the input ingredients including galaxy sample choice (with 
relevance to stellar mass profile and intrinsic galaxy shape), stellar IMF, 
stellar-to-halo mass relation, outer halo mass-concentration relation,
 velocity dispersion anisotropy profile,
central sub-kiloparsec stellar mass profile, and DM profile functional form.
 We investigate how the statistical results are varied in response to a 
variation of each ingredient. We find that all varied results have
similar curvature patterns in density slope profiles but have different values
of characteristic density slopes. We also find that when inputs are varied, 
the resulting distribution of velocity dispersion profile slope $\eta$ 
(equation~\ref{eq:lwvd}) is also changed. 

\begin{table*} % Table 2
\caption{ Various inputs and the resulting means of the luminosity-weighted
 LOSVD profile slope $\eta$ and the mass density profile absolute slope 
$\gamma_{\rm e}$ in the optical region ($0.1 R_{\rm eff}< r < R_{\rm eff}$). 
$s_{\eta}$ and $s_{\gamma_{\rm e}}$ are the resulting scatters of the sample.}
\begin{tabular}{cccccc}
\hline 
\hline 
 \# &  input &  $\langle\eta\rangle$  & $s_{\eta}$
   &  $\langle\gamma_{\rm e}\rangle$  & $s_{\gamma_{\rm e}}$  \\
\hline
\hline 
1 & standard (Table~1) & $-0.052$ & $0.028$ &  $2.124$  & $0.133$ \\
\hline 
2 & chi-square approach & $-0.059$ & $0.022$ &  $2.160$  & $0.111$ \\
\hline 
3 & \( \begin{array}{c} 
    2,000~C_r>2.86~{\rm random~ETGs} \\
    {\rm with~single~Sersic~fits} \end{array} \)
  &  $-0.052$  &  $0.034$  & $2.113$   &  $0.165$  \\
\hline 
4 & \( \begin{array}{c} 
       2,607~C_r>2.6~{\rm EGs~with}~\varepsilon<0.15 \\
      \& \log_{10}(R_{\rm eff}/R_{\rm bulge,eff}) < 0.19 \end{array} \)
  &  $-0.047$   & $0.035$  & $2.098$   & $0.157$ \\
\hline 
5 & \( \begin{array}{c} 
    2,000~C_r>2.6~{\rm random~ETGs} \\
   {\rm with~single~Sersic~fits} \end{array} \)
  &  $-0.044$ &  $0.040$  & $2.070$   &  $0.197$  \\
\hline 
 6 & Einasto DM model  & $-0.046$  &  $0.034$ &  $2.094$   & $0.163$ \\
\hline 
\( \begin{array}{c}  7 \\ 8  \end{array}\) &
  IMF \(\left\{\begin{array}{c}  {\rm SPS}\\ {\rm SPIDER}\end{array}\right.\)
 &    \(\begin{array}{c} -0.056\\ -0.046 \end{array}\) & 
   \(\begin{array}{c} 0.028  \\ 0.031  \end{array}\) &
  \(\begin{array}{c} 2.142 \\ 2.092  \end{array}\) &
   \(\begin{array}{c} 0.131 \\ 0.144  \end{array}\)  \\
\hline 
\( \begin{array}{c}  9 \\ 10  \end{array}\) &
$M_\star^{\rm Ch}$-$M_{200}$ \( \left\{\begin{array}{c} {\rm weak~lensing} \\ 
{\rm abundance~matching}  \end{array} \right.\) & 
  \(\begin{array}{c} -0.050 \\-0.058 \end{array}\) & 
   \(\begin{array}{c}  0.029 \\ 0.028  \end{array}\) &
  \(\begin{array}{c}  2.113  \\ 2.152  \end{array}\) &
   \(\begin{array}{c}  0.138  \\ 0.126  \end{array}\)  \\
\hline 
\( \begin{array}{c}  11 \\ 12  \end{array}\) &
$M_{200}$-$c_{\rm NFW}$ \( \left\{\begin{array}{c} 
30\%(\approx 2\sigma)~{\rm higher} \\ 
30\%(\approx 2\sigma)~{\rm lower}  \end{array} \right.\) & 
   \(\begin{array}{c} -0.045 \\ -0.060 \end{array}\) & 
   \(\begin{array}{c}  0.031 \\  0.027 \end{array}\) &
   \(\begin{array}{c} 2.085  \\ 2.170  \end{array}\) &
   \(\begin{array}{c} 0.144 \\  0.126   \end{array}\) \\
\hline 
 13 & radially varying $\beta(r)$ 
  &  $-0.051$  &  $0.055$  &  $2.131$ & $0.139$  \\
\hline 
 14 &$r_b/R_{\rm eff}=0.040\left(M_\star^{\rm Ch}/10^{11}{\rm M}_\odot\right)^{-0.55}$
 &  $-0.058$  &  $0.032$  & $2.123$    & $0.139$ \\
\hline 
 15 &  \( \begin{array}{c} 
 R_{\rm eff}~{\rm -dependent} \\
 \eta~{\rm(Fig.~\ref{VPRen})}~\&~\beta~{\rm(Fig.~\ref{bet3})} \end{array} \)
 &  $-0.054$  &  $0.028$  &  $2.125$  & $0.138$ \\
\hline
\hline 
\end{tabular}

\end{table*}

In Table~2 we list various input variations and the 
predicted values of $\langle\eta\rangle$ and $\langle\gamma_{\rm e}\rangle$
(population mean slopes for the radial range $0.1 R_{\rm eff} < r < R_{\rm eff}$).
The important points are as follows:

\begin{enumerate}
\item \emph{The spherical symmetry assumption and galaxy sample choice}: 
We have a large number of  galaxies selected by $C_r$, 
but empirical information for each galaxy is relatively limited compared with 
well-studied local ETGs (e.g.\ \citealt{Cap06,Tho,Cap13a}). Because of
 the limit of our data we have assumed the spherical symmetry. We have then 
selected nearly spherical galaxies ($\epsilon < 0.15$) to minimize the error 
arising from the spherical symmetry assumption. In reality, even nearly 
spherical galaxies have in some cases non-negligible disk components in their 
light distributions. Because the presence of non-negligible disk components can 
invalidate the spherical symmetry assumption, we have further narrowed our
selection by rejecting galaxies that have non-negligible disks. Our final sample
of 2,054 nearly spherical, disk-less and $C_r>2.86$ galaxies can be 
legitimately modelled relying on the spherical symmetry assumption. 

However, there still remain a couple of issues. 
First, as we are modelling nearly spherical and disk-less galaxies only, 
our results may not hold for the general population of ETGs.
For example, it is possible that intrinsic shapes of galaxies can increase
the intrinsic scatter of the mass profile. 

Second, all auxiliary empirical
properties of ETGs (e.g.\ the IMF distribution, the $M_\star-M_{200}$ relation, 
the distributions of $\beta$ and $\eta$) are for general populations of ETGs or
 red galaxies. If these properties are systematically different for spherical
galaxies, then our results will be biased. However, we do not find any
systematic difference between all ETGs and nearly spherical ETGs based on
the literature data (see, e.g., Fig.~\ref{VPdist} for the case of $\eta$).

As a way of estimating the systematic effects of non-spherical shapes 
we consider two samples of random ETGs. One sample is drawn randomly from
the $C_r > 2.86$ sample and the other sample from
a sample with a relaxed criterion of $C_r > 2.6$. 
We also consider a $C_r>2.6$ sample of nearly spherical and disk-less galaxies
to see the pure effects of changing the $C_r$ criterion.
 The results for the random $C_r > 2.86$ sample (fit \#~3 in Table~2) have 
almost the same mean values of the slopes $\eta$ and $\gamma_{\rm e}$ compared 
with the standard nearly spherical and disk-less sample (fit \#~1) 
but have larger scatters in good agreement with the observed scatters for 
random ETGs. The $C_r>2.6$ samples (fit \#~4 \& \#~5) have shallower mean
slopes and larger scatters. In particular, the predicted values of the mean VP 
slope $\langle\eta\rangle$ are marginally discrepant with the empirical value 
$\langle\eta\rangle=-0.058\pm 0.008$. Moreover, the random $C_r>2.6$ sample 
(fit \#~5) has larger scatter than observed. 

\item \emph{The assumption of parametric functional forms for DM distribution}: 
One of our underlying assumption is that the unknown DM distribution varies
smoothly in radius and can be modelled by parametric functional forms 
motivated by $N$-body simulations combined with baryonic effects. 
Because it is an empirical fact that the outer halo where baryonic effects are
negligible follows well the NFW functional form, we chose the gNFW model with
a generalized central slope as our standard model. To see possible systematic
errors of this assumption we also consider the Einasto model that is equally
well motivated. We have checked that contracted haloes due
 to baryonic effects produced by state-of-the-art cosmological hydrodynamic 
simulations can be well approximated by the gNFW or the Einasto profile, as can 
be checked numerically using the Contra code (\citealt{Gne,Gne11}). 
 This means that if baryon-induced halo contraction effects are real, 
 our models can mimic those effects. Also, note that our models can mimic 
effects of halo expansion as well if that occurs (for some galaxies). 
The result based on the the Einasto model (fit \#~6) has somewhat shallower
mean slopes compared with the standard result (fit \#~1) based on the gNFW 
model and interestingly larger scatter. Although the scatter of $\eta$ is 
more consistent with the observed scatter than the standard result, the mean 
$\langle\eta\rangle$ is marginally shallower than the observed mean.

\item \emph{Various empirical inputs}: We consider alternative choices of
various empirical inputs, fit \#~7 -- \#~14 in Table~2. 
The largest possible systematic differences are 
$\approx 0.006$-$0.008$ in $\langle\eta\rangle$
and $\approx 0.03$-$0.05$ in $\langle\gamma_{\rm e}\rangle$
 arising from the $M_\star^{\rm Ch}$-$M_{200}$ relation and the 
$M_{200}$-$c_{\rm NFW}$ relation. 
It is interesting to note that for the result using wildly varying VD
 anisotropies based on equation~\ref{eq:varbet} has unchanging mean slopes but
have almost twice larger scatter in $\eta$ ($s_\eta=0.055$ compared with 
$s_\eta=0.028$) while keeping the scatter in $\gamma_{\rm e}$ nearly unchanged.
This shows that anisotropy shapes can only affect the scatter of $\eta$ but
probably not the mean slopes. 

\item \emph{Correlations among empirical input parameters}: 
In our approach
each galaxy with measured parameter values ($\sigma_{\rm ap}$ which is on 
average closest to $\sigma_{\rm e2}$, $R_{\rm eff}$, $n$, $M_\star^{\rm Ch}$) 
is assigned various other parameter values 
[IMF variation with respect to the Chabrier IMF expressed in terms of 
$\delta_M = \log_{10}(M_\star/M_\star^{\rm Ch})$, halo mass $M_{200}$,
outer halo profile concentration expressed using the NFW profile $c_{\rm NFW}$, 
 anisotropy $\beta$ (and velocity dispersion profile slope $\eta$ for the
chi-square approach)]
using observed two-parameter relations including intrinsic scatter 
based on a wealth of empirical information. However, we ignored any possible
multiple correlations among input parameters.
 Here, we examine/discuss possible correlations and   
 investigate how the mass profile results are affected.

\begin{figure} %14
\begin{center}
\setlength{\unitlength}{1cm}
\begin{picture}(10,7)(0,0)
\put(-0.8,7.){\includegraphics{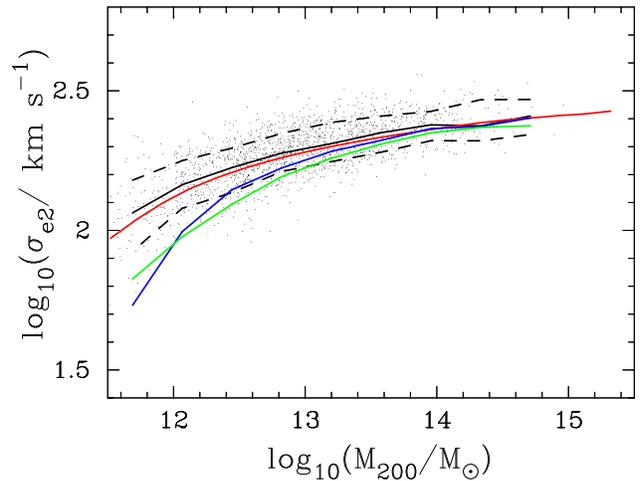}}
\end{picture} 
\caption{Data points and black curve show the posterior distribution of 2,054 
SDSS nearly spherical galaxies for the modelling result with the abundance 
matching $M_\star^{\rm Ch}$-$M_{200}$ relation (Table~2). (Dashed curves are 
$1\sigma$ scatters.) The red curve is the result of a bi-variate abundance 
matching of $M_\star^{\rm Ch}$ and $\sigma_{\rm e2}$ with $M_{200}$ by 
Chae et al.\ (2012). The difference
between the two is minor compared with the systematic differences with the
results (blue and green curves) based on different $M_\star^{\rm Ch}$-$M_{200}$ 
relations. Blue curve shows the result with standard $M_\star^{\rm Ch}$-$M_{200}$ 
relation and green curve with weak lensing relation (Table~2).
}
\label{MhV}
\end{center}
\end{figure}

First, when we assigned $M_{200}$ to each galaxy, 
we used the observed $M_\star^{\rm Ch}$-$M_{200}$ relation only.
In doing so we ignored any possible correlations, in particular 
with $\sigma_{\rm e2}$.  To see possible effects of the ignored correlation with
 $\sigma_{\rm e2}$ we compare in Fig.~\ref{MhV} our posterior 
$M_{200}$-$\sigma_{\rm e2}$ relation with the recent abundance matching output 
for which $M_\star^{\rm Ch}$ and $\sigma_{\rm e2}$\footnote{\cite{Chae12} used 
$\sigma_{\rm e8}$ which has been converted to $\sigma_{\rm e2}$ here.}
 were simultaneously assigned to $M_{\rm vir} \sim 1.2 \times M_{200}$ 
taking into account correlations (\citealt{Chae12}). 
Our posterior  $\sigma_{\rm e2}$-$M_{200}$ relation (data points and black line)  
deviates somewhat from the abundance matching relation (red line) in 
\cite{Chae12}, but the difference is certainly smaller than systematic 
differences due to the change of the $M_\star^{\rm Ch}$-$M_{200}$ relation.

Second, when we assigned $\delta_M$ and $c_{\rm NFW}$ we used empirical
relations respectively with $\sigma_{\rm e2}$ and $M_{200}$ only. Similarly to
the above case these assignments will have some systematic errors due to ignored
correlations with other parameters. However, again the systematic errors 
arising from uncertainties of the $\sigma_{\rm e2}$-$\delta_M$ and 
$M_{200}$-$c_{\rm NFW}$ relations themselves are likely to be dominating.

\begin{figure} %15
\begin{center}
\setlength{\unitlength}{1cm}
\begin{picture}(10,7)(0,0)
\put(-0.3,7.){\includegraphics{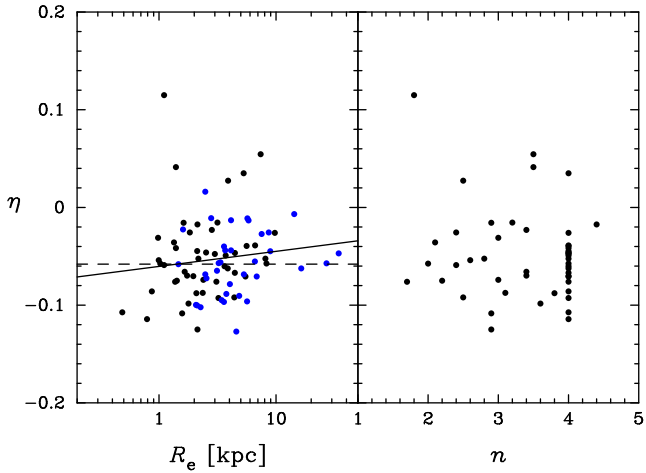}}
\end{picture} 
\caption{Distribution of $\eta$ against $R_{\rm eff}$ and $n$
for the galaxies shown in Fig.~4 except for $z \sim 0.25$ BCGs.
Values of $R_{\rm eff}$ come from de~Zeeuw et al.\ (2002) and 
Mehlert et al.\ (2000) respectively for 
black and blue data points. 
Values of $n$ shown on the right-hand side come from Cappellari et al.\ (2007).}
\label{VPRen}
\end{center}
\end{figure}

Finally, when we assigned $\beta$ (and $\eta$ for the chi-square approach) 
we assumed no correlations at all.
As shown in Fig.~\ref{VPdist}, $\eta$ has no correlation with $\sigma_{\rm e2}$
 for the ETGs that can match with our ETG samples.
Fig.~\ref{VPRen} shows $\eta$ against $R_{\rm eff}$ and $n$ using available data 
in the literature. There is possibly a weak trend with $R_{\rm eff}$ as shown by
 a solid line but no correlation is found with $n$. Fig.~\ref{bet3} shows 
$\beta$ against $\sigma_{\rm e2}$, $R_{\rm eff}$ and $n$ for 24 ETGs from 
\cite{Cap07}. There appears a trend with $R_{\rm eff}$ only but sample size
is too small to be meaningful. Fig.~\ref{etabeta} shows  $\eta$ against $\beta$
 and no correlation is found based on this small sample. To account for the 
effects of possible correlations of both $\beta$ and $\eta$ with $R_{\rm eff}$ 
we consider as an alternative choice the least-square fit relations shown in 
Fig.~\ref{VPRen} and Fig.~\ref{bet3}.

\begin{figure} %16
\begin{center}
\setlength{\unitlength}{1cm}
\begin{picture}(10,5)(0,0)
\put(-0.3,6.1){\includegraphics{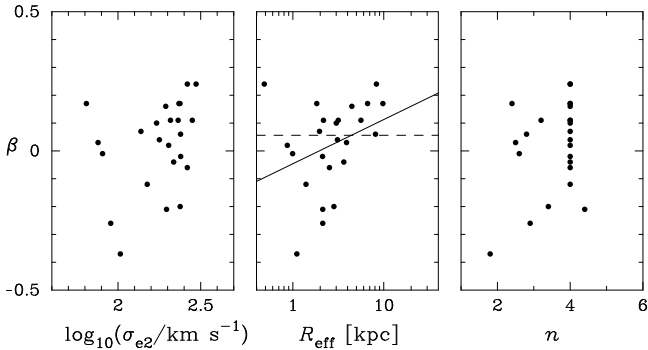}}
\end{picture}
\caption{Distribution of constant anisotropy $\beta$ for 24 nearby ETGs 
(a subsample of that shown in Fig.~4) modelled in  Cappellari et al.\ (2007).
There appears some trend with $R_{\rm eff}$ (solid line in the middle panel) 
based on this small sample. Values of $R_{\rm eff}$
and $n$ come respectively from de~Zeeuw et al.\ (2002) and  
Falc\'{o}n-Barroso et al.\ (2011).
}
\label{bet3}
\end{center}
\end{figure}

\begin{figure} %17
\begin{center}
\setlength{\unitlength}{1cm}
\begin{picture}(10,6)(0,0)
\put(-0.3,6.){\includegraphics{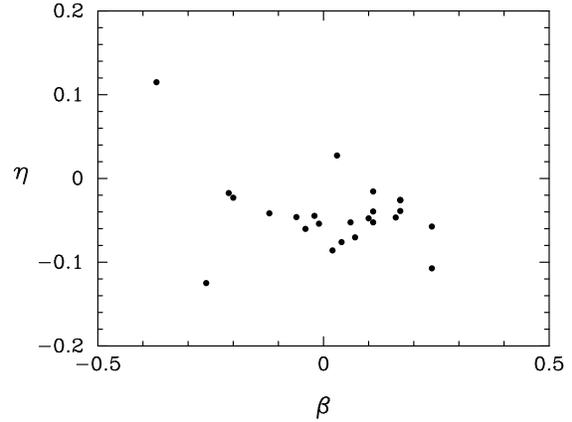}}
\end{picture}
\caption{No correlation is found between $\beta$ and $\eta$ 
for 24 galaxies shown in Fig.~16.}
\label{etabeta}
\end{center}
\end{figure}

\begin{figure} %18
\begin{center}
\setlength{\unitlength}{1cm}
\begin{picture}(8,9)(0,0)
\put(-0.3,-0.8){\includegraphics{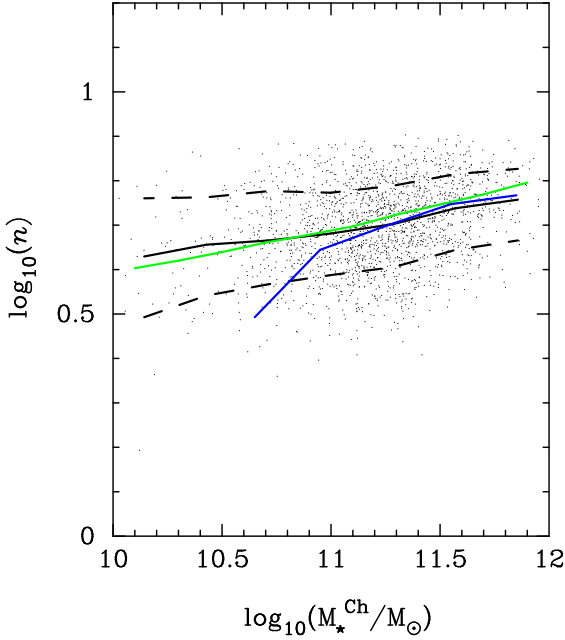}}
\end{picture}
\caption{Distribution of S\'{e}rsic index $n$ as a function of 
 $M_\star^{\rm Ch}$. Data points are 2,054 nearly spherical and disk-less
galaxies of our standard sample that are well-described by a single S\'{e}rsic 
function with the prior $n \le 8$.
Black curve shows the median value and dashed curves show $1\sigma$ scatters.
Green curve shows the median value for all 20,210 galaxies satisfying 
$C_r > 2.86$ with the prior $n \le 8$, 
not all of which can be well described by a
single S\'{e}rsic function. (This means that the distribution shown here for 
all ETGs is biased to some degree.) Blue curve is the median value for a 
published ETG sample based on a different selection by Guo et al.\ (2010). 
}
\label{nser}
\end{center}
\end{figure}

\end{enumerate}

\begin{figure} %19
\begin{center}
\setlength{\unitlength}{1cm}
\begin{picture}(9,7)(0,0)
\put(-0.7,7.3){\includegraphics{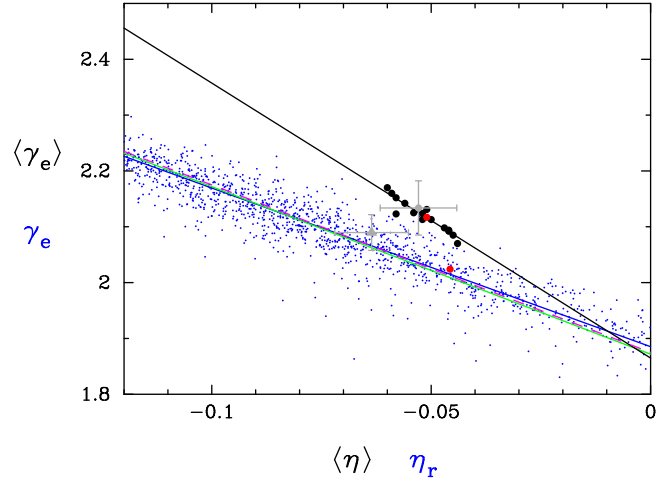}}
\end{picture} 
\caption{ Predicted relation between mass and velocity dispersion 
slopes within $R_{\rm eff}$ for our fiducial model. 
Black points represent mean values for
various modelling results obtained by varying model input ingredients.
 Two gray points with error bars represent the mean slopes for 9 normal 
ellipticals (filled circle) and 16 ETGs (filled triangle) of the Coma cluster
using the models constructed by Thomas et al.\ (2007).
Two red points represent the original and revised results based on a
semi-empirical approach (Chae et al.\ 2012). 
 Blue points represent individual slopes for the fiducial model.
  $\eta$ and $\eta_{\rm r}$ respectively denote
slopes for luminosity-weighted LOSVD (equation~11) and radial VD (equation~9).
Black and blue lines are least-square fit results.
The green line is the least-square fit result for the fixed slope of $-3$.
 The dashed purple line is the relation predicted by the Bertschinger (1985)
 spherical infall model and produced 
for pure DM haloes by $N$-body simulations (Taylor \& Navarro 2001).
}
\label{etagam}
\end{center}
\end{figure}

Fig.~\ref{etagam} shows the resulting distribution of 
$\langle\gamma_{\rm e}\rangle$ against $\langle\eta\rangle$ 
 for all 15 fits in Table~2.
There is a good correlation between the two parameters. The tight correlation
is remarkable considering that each result (or set of results) was obtained by 
varying a different ingredient. Caring little about all the details of the
inputs $\langle\gamma_{\rm e}\rangle$ is essentially dictated by 
 $\langle\eta\rangle$. This means that we can reliably estimate 
$\langle\gamma_{\rm e}\rangle$ just by measuring $\langle\eta\rangle$ for an
ETG sample without detailed dynamical modelling. 
 The least-square fit relation is given by 
$\langle\gamma_{\rm e}\rangle = a + b \langle\eta\rangle$ with
 $a= 1.865\pm  0.008$ and $b=-4.93\pm 0.15$.
From our considered 15 different fits
 $2.07 \lesssim \langle\gamma_{\rm e}\rangle\lesssim 2.17$ and
 $-0.060 \lesssim \langle\eta\rangle \lesssim -0.044$ along with
$0.22 \lesssim s_{\eta} \lesssim 0.55$ and 
$0.11 \lesssim s_{\gamma_{\rm e}} \lesssim 0.20$.  
 
 With our measured $\langle\eta\rangle=-0.058\pm 0.008$ from section~2.6,  
we have $\langle\gamma_{\rm e}\rangle=2.15\pm 0.04$. The value
$\langle\eta\rangle=-0.066$ measured by \cite{Cap06} corresponds to
$\langle\gamma_{\rm e}\rangle \approx 2.18$. Hence our results suggest
a super-isothermal mean profile in the optical region with an intrinsic scatter
 of $0.1 \lesssim s_{\gamma_{\rm e}} \lesssim 0.2$. 
\cite{Bol} have recently carried out stellar dynamics plus 
strong lensing analyses of dozens of ETGs at $z \la 0.6$. For 14 ETGs from
\cite{Bol} at a redshift range $0.05 < z < 0.15$ similar to our sample, we
find $\langle \gamma_{\rm e} \rangle = 2.242 \pm 0.034$ with
 $\sigma_{\gamma_{\rm e}}=0.121$. This result is even more steeper
but consistent at a $2\sigma$ level with our results.
Our results are somewhat ($\sim 2\sigma$) steeper than the mean profile of 
16 Coma cluster ETGs, $\langle \gamma_{\rm e} \rangle = 2.064 \pm 0.042$ with 
$\sigma_{\gamma_{\rm e}}=0.163$ (\citealt{Tho}). 
 However, if we consider only 9 normal ellipticals excluding lenticular 
and cD/D galaxies from the Coma cluster, the mean profile is 
$\langle \gamma_{\rm e} \rangle = 2.134 \pm 0.049$ with 
$\sigma_{\gamma_{\rm e}}=0.137$ in closer agreement with our results.
Our results are also somewhat steeper than, but consistent at a $2\sigma$ level
with, the SLACS results (e.g.\ \citealt{Aug,Bar}): constant slope 
$\gamma=2.074^{+0.043}_{-0.041}$ with $s_\gamma=0.144^{+0.055}_{-0.014}$ 
(\citealt{Bar}). However, unlike the \cite{Bol} result quoted above this result
is for redshift range of $0.06\la z \la 0.5$. As is discussed by \cite{Bol}
and in section~8 below, the inferred $\gamma$ may depend on the redshift
range, so that caution should be taken in comparing the SLACS results with 
our result.

\section{Comparison with previous dynamical analyses results based on SDSS aperture velocity dispersions and photometric data}

SDSS aperture velocity dispersions and photometric data have been widely used
in the literature to infer dynamical masses and dark matter fractions in the
optical region (e.g.\ \citealt{Pad,Tor12}) and 
to test halo contraction/expansion models and stellar IMFs 
(e.g.\ \citealt{Sch,Dut11,Dut13,DT13}). 
It has been found that the SDSS data require non-universal IMFs and/or
non-universal halo response to baryonic physics
(\citealt{Dut11,Dut13,DT13}). However, none of these studies
have tried to predict mass profiles or considered velocity dispersion 
profiles, although \cite{DT13} have used the SLACS mass profile slope 
distribution ($\langle\gamma\rangle\approx 2.08$,
$s_\gamma\approx 0.16$; \citealt{Aug}) to further study IMFs and halo response. 
\cite{DT13} argue that for massive ellipticals with mean
$\log_{10}(\sigma_{\rm e2}/{\rm km}~{\rm s}^{-1})=2.40$ similar to the SLACS 
systems, uncontracted NFW haloes combined with the Salpeter IMF can reproduce 
the SLACS slope distribution. However, their argument depends on 
two assumptions that the SLACS slope distribution at higher
redshifts can match low-$z$ SDSS galaxies and the
fiducial stellar mass based on the Chabrier IMF is correct (remember that
$M_\star^{\rm Ch}$ itself is uncertain as mentioned in section~2.2). Nevertheless,
the \cite{DT13} result differs by just $\sim 2\sigma$ from our best estimate
and is within our systematic errors (cf.\ Table~2).
In particular, it appears that if one fully considers realistic systematic
errors halo contraction effects are inconclusive yet largely due to 
uncertainties in stellar masses involving IMFs (in preparation). 

 Recently a Jeans analysis of ETGs has been carried out by \cite{Chae12} 
based on abundance matching between SDSS ETGs and the Bolshoi $N$-body 
simulation haloes (\citealt{Kly}) and previously available statistical empirical
 properties of ETGs. The focus of that analysis was to compare the resulting 
DM profile (baryonic effects included) with the DM-only $N$-body 
simulation profile.  
A by-product was that mass profile slope values measured at $R_{\rm eff}/2$ 
were on average close to isothermal. This is in contradiction with our 
steeper-than-isothermal profiles presented above. Specifically, for the case of
using the same abundance matching $M_\star^{\rm Ch}$-$M_{200}$ relation our new 
result $\langle\gamma_{\rm e}\rangle\approx 2.15$ (fit \#~10, Table~2) 
is steeper than $\langle\gamma_{\rm e}\rangle\approx 2.02$ 
(the lower red point in Fig.~\ref{etagam})
 for a total of 3,000 ETGs (1,000 ETGs respectively of 
$M_{200}=10^{12.5}{\rm M}_\odot$, $10^{13}{\rm M}_\odot$ and $10^{13.5}{\rm M}_\odot$).
Several factors can account for this difference.

First of all, as shown in Fig.~\ref{nser} our newly measured values of 
S\'{e}rsic index $n$ are higher than the previously adopted values 
(\citealt{Guo}) for $M_\star^{\rm Ch} \la 10^{11}{\rm M}_\odot$. 
This is in part due to the difference in sample definitions. The \cite{Guo} 
sample includes low concentration spheroidal galaxies at low 
stellar masses while our sample includes only high concentration 
(i.e.\ $C_r > 2.86$) galaxies.  When the semi-empirical approach
is carried out with our new distribution of $n$ for all ETGs (green curve) and
the ATLAS$^{\rm 3D}$ IMF distribution we have a higher 
$\langle\gamma_{\rm e}\rangle\approx 2.12$ 
(the upper red point in Fig.~\ref{etagam}).

The remaining difference of $\approx 0.03$ can be attributed to
other factors including differences in the posterior distribution of
parameters (in particular $R_{\rm eff}$) and their correlations owing to
the different modelling procedure. In particular, in the semi-empirical approach
the posterior distribution of $\eta$ always had a mean shallower than $-0.058$
and was tried to approximately match $\langle\eta\rangle=-0.053$. 
Fig.~\ref{etagam} shows that the revised semi-empirical result falls right 
 on the empirical $\langle\eta\rangle$--$\langle\gamma_{\rm e}\rangle$ line while
 the original result deviates somewhat. In other words,
whatever factors contributed to the difference in $\langle\gamma_{\rm e}\rangle$
it can be explained by the difference in the posterior value of 
$\langle\eta\rangle$.

Despite the large difference in $\langle\gamma_{\rm e}\rangle$ we find that 
total mass profiles from the semi-empirical analysis show
similar curvature in the mean density slope profiles.

\section{Implications for galaxy formation: universal pseudo-phase-space density profile?}

The radial systematic pattern of each mass profile (Fig.~\ref{rhogamex}, 
Fig.~\ref{rhogam}) and the systematic variation of the characteristic slopes 
(Fig.~\ref{gamvar}) as a function of some parameters, in 
particular the size and the projected effective surface density, implies that 
there is no natural attractor or strict conspiracy
for total mass profiles of ETGs. The isothermal model is good only as a 
constant-slope approximation to the mean inner halo of the ETG population. 
ETGs are complex systems of luminous and dark matter and 
the interplay between them may create super-isothermal and sub-isothermal 
regions. 
When predictions of galaxy formation theories are compared with observed 
galaxies, the same radial ranges should be used and more importantly radially 
varying slopes  may need to be considered.

The tight correlation between total mass and stellar VD slopes shown in 
Fig.~\ref{etagam} is reminiscent of the pseudo phase-space density profile
for DM distribution produced by $N$-body simulations (\citealt{TN}). A universal
 power-law behaviour of the pseudo-phase-space density of DM (\citealt{TN}) 
given by 
$\rho_{\rm DM}(r)/\sigma_{\rm DM}^3(r) \propto r^{-\nu}$ ($\nu\approx 1.875$) 
implies $\gamma_{\rm DM}=\nu - 3 \eta_{\rm DM}$ where $\sigma_{\rm DM}(r)$ is 
the VD in the 3-dimensional physical space, 
$\gamma_{\rm DM}=-d\ln\rho_{\rm DM}(r)/d\ln r$, 
and $\eta_{\rm DM}=d\ln\sigma_{\rm DM}(r)/d\ln r$. 
A different slope of $\approx -4.93$ in our relation 
$\langle\gamma_{\rm e}\rangle \approx 1.86 -4.93 \langle\eta\rangle$ can be 
attributed to the fact that $\langle\eta\rangle$ is for the LOSVD. Indeed,
if we consider radial stellar VD $\sigma_{\rm r}(r)$ 
(cf.\ equation~\ref{eq:jeans}) we recover a slope of $\approx -3$. 
Fig.~\ref{etagam} shows distributions of  $\gamma_{\rm e}$ (total mass
absolute slope within $R_{\rm eff}$) 
and $\eta_{\rm r}$ ($\sigma_{\rm r}(r)$ slope within 
$R_{\rm eff}$) for our fiducial model. 
There is some intrinsic scatter but the least-square fit
linear relation is $\gamma_{\rm e}\approx 1.89 -2.84 \eta_{\rm r}$ 
and $\gamma_{\rm e}\approx 1.87 -3 \eta_{\rm r}$ for the fixed slope of
$-3$ which are very similar to the above relation for DM.
This implies that there may exist a universal profile of pseudo phase-space
density-like quantities in any collisionless components in dynamical 
equilibria (\citealt{Chae13}).

\section{Implications for interpretation of gravitational lensing
analyses}

Radially varying mass profiles of ETGs call for a careful interpretation of
gravitational lensing, in particular strong lensing. Strong 
lensing depends on the projected 2-dimensional density slope $\gamma_{\rm 2D}$
 at the Einstein radius $R_{\rm Ein}$. 
The redshift evolution of $R_{\rm Ein}$ in conjunction with the
radial variation of the density slope (Fig.~\ref{rhogam}) can produce an 
apparent redshift evolution of a mass profile from an analysis of a 
strong lensing sample. 
Let us make a quantitative prediction based on a strong lens sample recently 
analysed by \cite{Bol}. 
The observed mean ratio $\langle R_{\rm Ein}/R_{\rm eff}\rangle$
 evolves from $\approx 0.5$ at $z=0.1$ to $\approx 1$ at $z=0.55$ with a slope
$d\log \langle R_{\rm Ein}/R_{\rm eff}\rangle/dz = 0.65\pm 0.12$ while 
$\langle R_{\rm eff}\rangle$ remains nearly unchanged (\citealt{Bol}). 
According to our results for ETG mass profiles
 the mean density slope $\langle \gamma_{\rm 2D}\rangle$ varies as 
$d\langle\gamma_{\rm 2D}\rangle/d\log r = -0.45 ^{+0.14}_{-0.03}$ between 
$r=0.5 R_{\rm eff}$ and $R_{\rm eff}$ (Fig.~\ref{rhogam}), which gives 
$d\langle\gamma_{\rm 2D}\rangle/dz = -0.29 ^{+0.18}_{-0.12}$.
Suppose we are using a constant-slope model for the lensing galaxy to 
fit the lensed image geometry and magnification ratios, then we would get
$d\langle\gamma\rangle/dz =-0.29^{+0.18}_{-0.12} $ (note that for the constant
slope profile $\gamma=\gamma_{\rm 2D}+ 1$).
The magnitude of this evolution is  consistent (within 
$\sim 2\sigma$) with the reported evolution 
$d\langle\gamma\rangle/dz=-0.60\pm 0.15$ from a combined analysis of 
strong lensing and stellar dynamics using a constant-slope model 
(\citealt{Bol}). 
 It is unclear whether this means the reported evolution is an artefact 
due to the radially varying slope or real because it was also based on 
stellar dynamics. However,  our results indicate that it is possible 
 that the reported evolution is due, at 
least in part, to the radially varying slope.

The density slopes at $R_{\rm eff}$ we derive are close to isothermal for
massive ellipticals, though super-isothermal within $R_{\rm eff}$, and
are consistent with lens modelling results (\citealt{RKK}) for intermediate 
redshift ($0.3 \lesssim z \lesssim 1$) strong lensing galaxies whose Einstein 
radii are on average close to $R_{\rm eff}$.

\section{Conclusions}

 Assembling statistically significant numbers of mass models of SDSS 
concentration-selected elliptical galaxies we find the following:
\begin{itemize}
\item Two-component mass models reproduce simultaneously the aperture velocity
dispersion and the observed velocity dispersion profile slope distribution,
but single-component mass models with constant or monotonically varying slope 
cannot successfully predict the observed velocity dispersion profile slope 
distribution without biasing the aperture velocity dispersion.
\item For the region within $R_{\rm eff}$
 there is a tight correlation between the mass density profile mean absolute 
slope $\langle\gamma_{\rm e}\rangle$ and the velocity dispersion profile mean 
slope $\langle\eta\rangle$ as 
$\langle\gamma_{\rm e}\rangle=(1.865\pm 0.008)+(-4.93\pm 0.15)\langle\eta\rangle$
 with little sensitivity to the details of the inputs and modelling procedures.
From this relation one can estimate $\langle\gamma_{\rm e}\rangle$ from
a measurement of $\langle\eta\rangle$ without detailed dynamical modelling.
\item  The tight correlation between the density and VD profile slopes 
implies an approximately universal pseudo phase-space density power-law profile
$Q(r)\equiv \rho(r)/\sigma_{\rm r}^3(r)\propto r^\nu$ with $\nu \approx -1.87$, 
where $\rho(r)$ is the total mass profile and $\sigma_{\rm r}(r)$ is the radial 
stellar VD. 
\item The radial variation of the total density slope we find in 
this study may result in an apparent redshift evolution of mass profile in 
strong lensing observations. 
\end{itemize}

The findings presented in this paper can provide 
interesting constraints on galaxy formation models. 
In future studies it would be 
interesting to compare  density and velocity dispersion profiles for galaxies 
forming in the concordance $\Lambda$CDM model and test whether they exhibit 
radial variation and correlation of slopes similar to those found in this study
 for observed galaxies. 

\bigskip

 We are grateful to Michele Cappellari, Charlie Conroy and Surhud More for
 providing their published and unpublished empirical results as tables. 
We thank them and Drew Newman for assistance in understanding their 
empirical results. We also thank Joshua Frieman for past collaboration and 
useful comments on an early draft. We gratefully acknowledge the anonymous 
referee's comments that helped us improve the manuscript significantly.
MB acknowledges support from NASA grant ADP/NNX09AD02G.
AVK acknowledges support by NSF and NASA via grants 
OCI-0904482 and NNX13AG81G and by the Kavli Institute for Cosmological Physics 
at the University of Chicago through grants NSF PHY-0551142 and PHY-1125897 
and an endowment from the Kavli Foundation and its founder Fred Kavli.

\bibliographystyle{mn2e}

\end{document}